\documentclass[
reprint,
superscriptaddress,
showkeys,
longbibliography,
frontmatterverbose, 
nofootinbib,
 amsmath,amssymb,
 aps,
]{revtex4-2}

\usepackage{graphicx}% Include figure files
\usepackage{dcolumn}% Align table columns on decimal point
\usepackage{bm}% bold math
\newcommand{\Ca}{\mathrm{Ca}}

\begin{document}

\preprint{APS/123-QED}

\title{Wetting on Silicone Surfaces\\ 
%Wetting on polydimethylsiloxane (PDMS) coatings: coating structures, contact mechanics and drop sliding\\
%Statics and Dynamics of Wetting on Silicone Surfaces\\ 
%Wetting on Silicone Surfaces: Liquid-Infused Surfaces, Elastomers, and Liquid-Like Surfaces\\
%Static and Dynamic Wetting of Soft, Lubricated, and Liquid-Like Silicone Surfaces
%\color{red}{please add suggestions}\\
}% Force line breaks with \\

\author{Lukas Hauer}
\email{lukas.hauer@hu-berlin.de}
\affiliation{Institute for Biology, Humboldt-Universität zu Berlin, 10115 Berlin, Germany}%Lines break automatically or can be forced with \\
\affiliation{Physics at Interfaces, Max Planck Institute for Polymer Research, Ackermannweg 10, 55128 Mainz, Germany}
 
\author{Abhinav Naga}
 \affiliation{Department of Physics, Durham University, DH1 3LE, United Kingdom}

\author{Rodrique G. M. Badr}
\affiliation{Institut für Physik, Johannes Gutenberg-Universität Mainz, Staudingerweg 7-9, 55099 Mainz, Germany}
%I would go for the English name

\author{Jonathan T. Pham}
 \affiliation{Department of Chemical and Environmental Engineering, University of Cincinnati, Cincinnati, 45221 OH, USA}%Lines break

\author{William S. Y. Wong}
 \affiliation{Department of Applied Physics, School of Science, Aalto University, 02150 Espoo, Finland}%Lines break
 
\author{Doris Vollmer}
\affiliation{Physics at Interfaces, Max Planck Institute for Polymer Research, Ackermannweg 10, 55128 Mainz, Germany}

\date{\today}% It is always \today, today,

\begin{abstract}
Silicone is frequently used as a model system to investigate and tune wetting on soft materials. Silicone is biocompatible and shows excellent thermal, chemical, and UV stability. Moreover, the mechanical properties of the surface can be easily varied by several orders of magnitude in a controlled manner. Polydimethylsiloxane (PDMS) is a popular choice for coating applications such as lubrication, self-cleaning, and drag reduction, facilitated by low surface energy. Aiming to understand the underlying interactions and forces, motivated numerous and detailed investigations of the static and dynamic wetting behavior of drops on PDMS-based surfaces. Here, we recognize the three most prevalent PDMS surface variants, namely liquid-infused (SLIPS/LIS), elastomeric, and liquid-like (SOCAL) surfaces. To understand, optimize, and tune the wetting properties of these PDMS surfaces, we review and compare their similarities and differences by discussing (i) the chemical and molecular structure, and (ii) the static and dynamic wetting behavior. We also provide (iii) an overview of methods and techniques to characterize PDMS-based surfaces and their wetting behavior. The static and dynamic wetting ridge is given particular attention, as it dominates energy dissipation, adhesion, and friction of sliding drops and influences the durability of the surfaces. We also discuss special features such as cloaking and wetting-induced phase separation. Key challenges and opportunities of these three surface variants are outlined. 

\end{abstract}

%Use showkeys class option if keyword
\maketitle

%\tableofcontents

\section{Introduction}

Silicones are abundant materials with global production margins exceeding a million tonnes annually. As a surface coating, the material is widely used in numerous industries, ranging from the medical and energy sectors to the personal care and automotive/aerospace industries. The low surface energy of silicones promotes low friction and adhesion with contacting solids and liquids, providing desirable properties for self-cleaning, lubrication, anti-icing/-biofouling, drag-reduction, and enhanced heat and mass transfer \cite{bhowmick_current_2008,griffith_polymeric_2000,yoda_elastomers_1998,lim_polymer_2010,greenham_semiconductor_1996,lalia_review_2013,eduok_recent_2017,golovin_low_2019,hu_siliconebased_2020,duffy_rapid_1998}. The most common silicone is polydimethylsiloxane (PDMS). PDMS is biocompatible, water repellent, and flexible while being chemically/thermally stable \cite{maitz_applications_2015,staapolonia_study_2019,moretto_silicones_2000,butts_silicones_2002,chung_skininterfaced_2020,kim_soft_2020,xia_soft_1998,ozbolat_3d_2018,morbioli_practical_2020}. The chemical inertness, low miscibility in water, and non-toxicity make PDMS mostly environmentally harmless \cite{fendinger_polydimethylsiloxane_2005}. Recent bacterial studies showed a biodegradability of the polymer \cite{sarai_directed_2024}, enabling a sustainable product life cycle. The mechanical properties of PDMS can be tuned over many orders of magnitude and the synthesis is safe and mostly straightforward when using commercially available two-part systems, requiring no advanced training. Thus, PDMS is readily used in labs as a model system for soft materials.\\
Due to the chemical and mechanical properties, PDMS is used as a surface coating: The softness and flexibility of PDMS yield excellent lubrication properties \cite{klein_forces_1991,klein_shear_1996,kreer_polymerbrush_2016} which can be utilized for joint lubrication \cite{klein_repair_2009} or drag reduction \cite{solomon_drag_2014,lee_interfacial_2014}. The low (lateral) adhesion is exploited for applications in anti-icing \cite{subramanyam_ice_2013,golovin_low_2019,kim_liquidinfused_2012,wong_capillary_2020}, anti-marine fouling \cite{amini_preventing_2017}, and anti-microbial formation \cite{epstein_liquidinfused_2012,leslie_bioinspired_2014,chen_immobilized_2017,lavielle_lubrication_2021,armugam_broad_2021}. Heat and mass transfer can be optimized as PDMS surfaces provide high nucleation rates while maintaining drop mobility \cite{sokuler_softer_2010,xiao_immersion_2013,cho_nanoengineered_2017}. PDMS is also used to interface biological systems, e.g., to control the growth and the size of cells \cite{lee_compatibility_2004,sotiri_tunability_2018}.\\
PDMS-based surfaces come in different forms with various material and wetting properties \cite{owen_silicone_2021}. Variations in polymer chain configuration give a broad design space for PDMS-based surfaces. The chain-chain interactions (cohesion and friction) vary with the tunable chain length or with the addition of covalent cross-links. The adhesive interactions with the underlying substrate can be physical (e.g., capillary forces) or chemical (e.g., covalent bonding).\\

To provide optimal surface functionality, understanding the wetting interactions in a given application is essential. Misinterpretation of these interactions can lead to misuse, malfunction, and deterioration of the surface. However, such understanding is challenging as the interactions are complex with strong coupling between the different components, and require consideration on many levels. During wetting, various forms of mass, momentum, and energy exchange take place on length scales ranging from the molecular structure of the polymer (\AA) to the macroscopic drop ($\mathrm{cm}$) \cite{butt_adaptive_2018,daniel_probing_2023, hartmann_drops_2024}. The associated timescales vary from mere picoseconds up to several days. The liquid drop can mix to some extent with the underlying polymer \cite{wong_adaptive_2020,schubotz_memory_2021,li_adaptation_2021}. The sub-components that make up the polymer matrix of solid PDMS can undergo chemical reactions or phase changes. This alters the molecular structure of either or both the surface and the liquid \cite{kleingartner_exploring_2013,butt_adaptive_2018}. Liquid PDMS chains (e.g., oil) can engulf and cloak the drop \cite{bergeron_monolayer_1996,mahadevan_fourphase_2002,kreder_film_2018}. The cloak changes the drop affinity to the surface and therefore affects the wettability. Interactions between the drop and surface are particularly high near the so-called “three-phase-contact line” \cite{berry_simple_1974,herring_simulation_2010}. At this location, the drop meets with the surface and the surrounding medium (a third fluid such as air). Here, the surface tension of the sessile drop exerts pulling stress on the surface. On many rigid surfaces such as metals or glass, the pulling stress is minor compared to the internal material stresses. PDMS, however, is soft, even when it comes as a rubber. Hence, the surface tension-induced pulling stress deforms the surfaces to an annular “wetting ridge” \cite{carre_viscoelastic_1996,clanet_onset_2002,park_selfspreading_2017,bico_elastocapillarity_2018}. On very soft surfaces the wetting ridge grows large enough that it becomes visible by the bare eye.\\

\begin{figure*}
    \centering
    \includegraphics[width=1\textwidth]{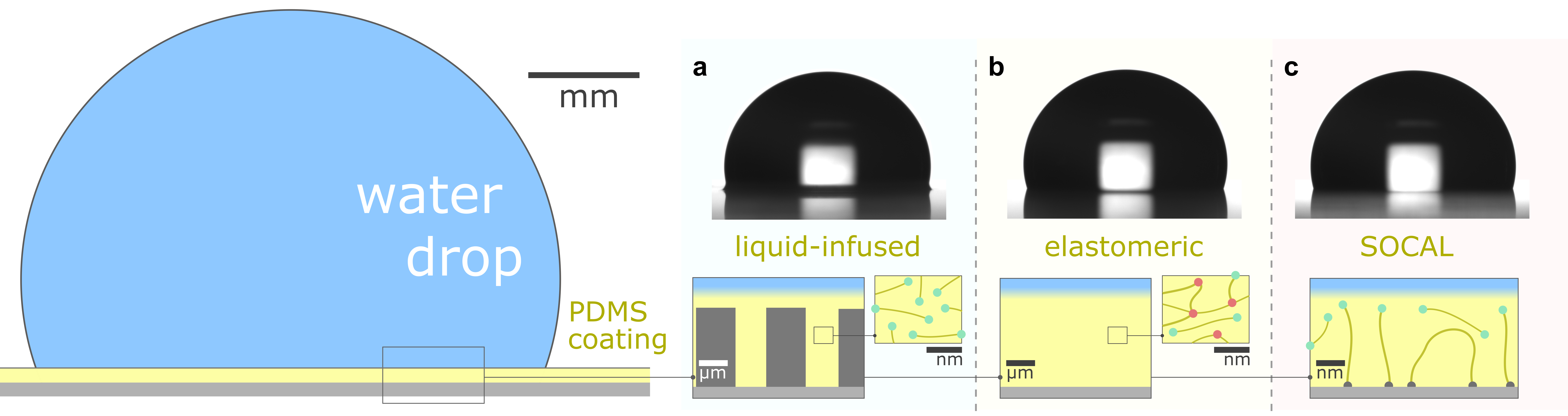}
    \caption{\label{fig:pdms_based_coatings} PDMS surface with sessile water drop. Scale bars serve as order of magnitude guide. Depending on their molecular chain configuration and substrate arrangement, PDMS surfaces can be a) liquid-infused, b) elastomeric, or c) molecularly attached (SOCAL) to a rigid substrate. The thickness of the surface varies from mm and $\mu$m (LIS, elastomeric) to nm (in particular SOCAL). Top row shows shadowgraphs of sessile drops ($10~\mathrm{\mu l}$) on each PDMS surface type, each exhibiting $\theta>90^\circ$. a) Low molecular weight chains with free ends (green points) behave as a Newtonian liquid, obtaining viscosities between $1~\mathrm{mPa~s}\leq\eta \leq 10~\mathrm{kPa~s}$. Rough/porous solid substrate texture provides support for oil lubrication and increases surface retention. b) Crosslinks (red points) create a network mesh with elasticity. Free chains (PDMS oil) can reside inside the network and rearrange freely. The degree of crosslinking and the amount of free chains inside the network determine the softness and lubrication of elastomer surfaces. c) PDMS chains are covalently bonded to anchor sites (gray points). The free ends remain flexible and maintain liquid-like surface properties on SOCAL surfaces. Few chains are grafted on both chain ends to the surface. Illustration and shadowgraphs adapted from \cite{hauer_wetting_2023}.}
\end{figure*}

This review aims to present, discuss, and compare the physicochemical properties of variations of PDMS surfaces during static and dynamic wetting. We first introduce the chemical and structural features of PDMS and PDMS-based surfaces. We focus on the three most frequently used PDMS-based surfaces: i) \textit{liquid infused surfaces } (cf. lubricant infused surface - LIS \cite{lafuma_slippery_2011}/slippery lubricant infused porous surface - SLIPS \cite{wong_bioinspired_2011}), where liquid PDMS oil impregnates a porous substrate structure, Fig. \ref{fig:pdms_based_coatings}a. ii) \textit{elastomeric surfaces} (cf. silicone gel surface), where PDMS chains are cross-linked to give structural integrity, Fig. \ref{fig:pdms_based_coatings}b. iii) \textit{slippery omniphobic covalently attached liquid-like surfaces} (cf. SOCAL, liquid-like surfaces  \cite{milner_theory_1988,buddingh_liquid_2021,daniel_origins_2018,chen_omniphobic_2023}), where PDMS chains form a nanometric thin surface layer, Fig. \ref{fig:pdms_based_coatings}c.\\
After establishing the structural differences in chain-chain and chain-surface organization on the different surfaces, we discuss their characteristic wetting behavior. We consider static and dynamic wetting - with a particular focus on the wetting ridge. This feature is ubiquitous amongst PDMS-based surfaces and the most pivotal element for wetting \cite{shanahan_viscoelastic_1995,gupta_numerical_2023,semprebon_apparent_2021,jerison_deformation_2011,pericet-camara_solidsupported_2009,long_static_1996,tress_shape_2017}, drop mobility \cite{long_static_1996,sadullah_drop_2018,karpitschka_droplets_2015,shanahan_viscoelastic_1995,keiser_drop_2017,zhao_geometrical_2018,khattak_direct_2022}, and surface durability \cite{hourlier-fargette_extraction_2018,hourlier-fargette_role_2017,jensen_wetting_2015,jeon_moving_2023,cai_fluid_2021,wong_adaptive_2020,hauer_phase_2023}. Contact friction mostly dissipates in the ridge \cite{daniel_oleoplaning_2017,keiser_drop_2017,karpitschka_droplets_2015,keiser_universality_2020,roche_complexity_2024}. Therefore it is critical for lubrication and drop sliding. However, it also accumulates surface material, and contacting objects may entrain surface material in the wetting ridge \cite{tran_wetting_2023}. Eventually, this can cause surface deterioration and malfunction. We discuss secondary effects such as PDMS-drop cloaking \cite{naga_how_2021,badr_cloaking_2022,sharma_enhanced_2022} which influences the wetting behavior and surface durability.\\
We emphasize the importance of the surface-associated wetting properties as they are directly related to \textit{lubrication} and \textit{durability}: These are the two main concepts for optimized surface functionality.
This perspective will help to gain a generalized, fundamental understanding of these most important PDMS-based surface variants, which helps to create guidelines for optimized and long-lasting PDMS surfaces. 

\section{PDMS-Based Surface - Structure and Properties}

\begin{figure}[h!]
    %\centering
    \includegraphics[width=0.5\textwidth]{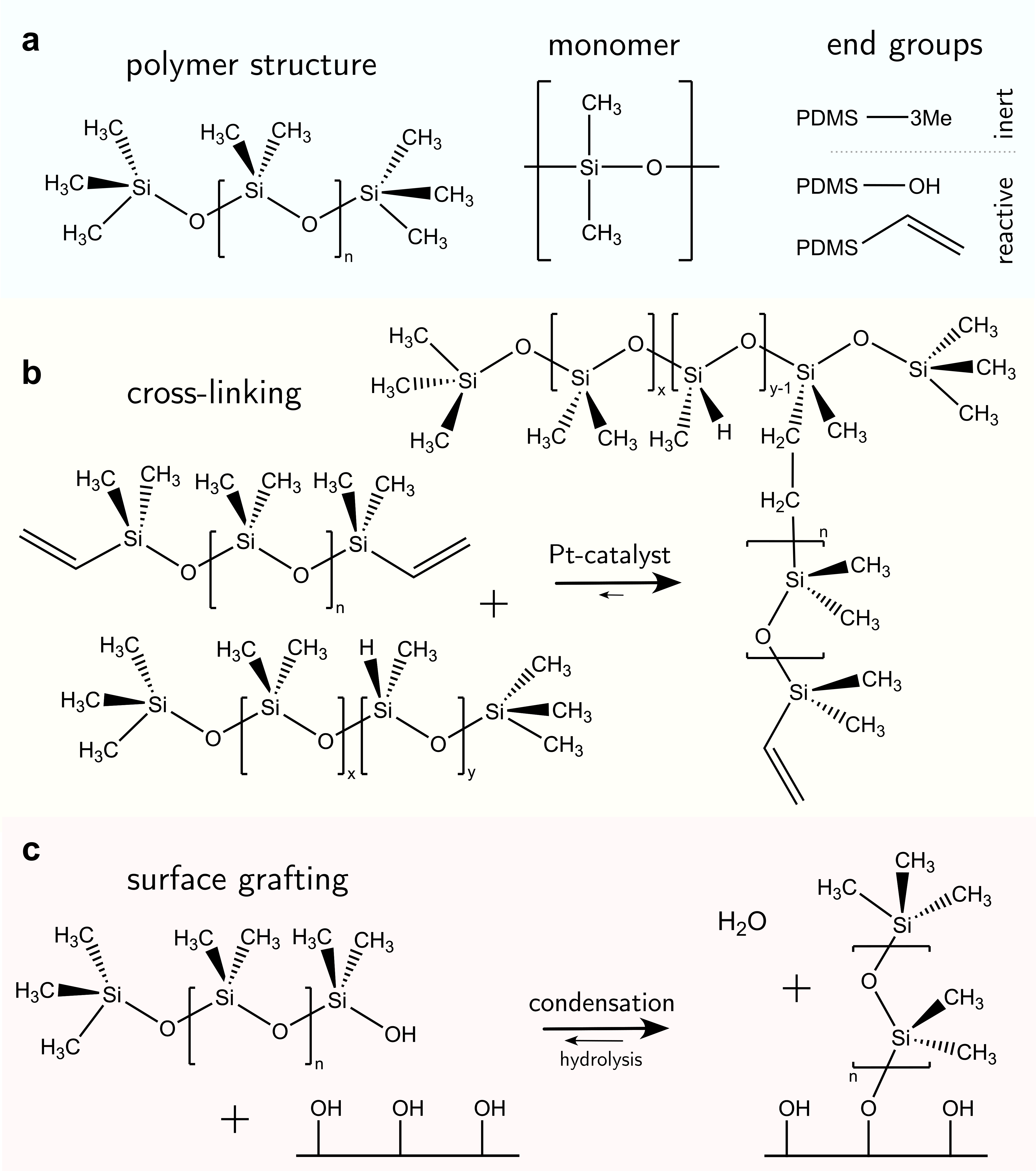}
    \caption{\label{fig:pdms_strucutre}Chemical structure of polydimethylsiloxane. a) PDMS can be oligomeric ($n < 100$) with low viscosity ($1-5~\mathrm{mPa~s}$) or assemble into long chains with high viscosities (up to $\eta=10~\mathrm{kPa~s}$). The silicon-oxygen backbone provides chain flexibility while the methyl side chains yield water repellency. The end groups can be inert (trimethylsilyl) or reactive (hydroxyl, vinyl). b) Hydrosilylation reaction between linear vynil-terminated PDMS chains (top) and cross-linking agents with methylhydrosiloxane monomers create branches and networked elastomeric polymers. Here, the cross-linking agent is a polymer chain with $x$ dimethylsiloxane and $y$ methylhydrosiloxane monomers that are randomly arranged. The network branches are randomly distributed on the cross-linker chain and terminal on the PDMS base chain. c) Hydroxyl-terminated PDMS chains can be grafted on surface hydroxyl groups that naturally occur on metals/metalloid oxides, or can be functionalized with hydrogen plasma.}
\end{figure}

PDMS-based surfaces come in various configurations with associated material properties and wetting behaviors. We consider three common types of PDMS-based surfaces, i) PDMS oil-based surfaces (e.g., liquid-infused surfaces), Fig. \ref{fig:pdms_based_coatings}a, ii) PDMS elastomeric surfaces (e.g., crosslinked polymers), Fig. \ref{fig:pdms_based_coatings}b, and iii) PDMS SOCAL surfaces (e.g., liquid-like polymer chains), Fig. \ref{fig:pdms_based_coatings}c. Each PDMS-based surface has the same monomeric building block, and consequently, some intrinsic properties are shared among the surface types.\\
The monomer of PDMS has an inorganic siloxane backbone and two methyl side chains, Fig. \ref{fig:pdms_strucutre}a center. The silicon-based backbone sets it apart from alkyl (hydrocarbon) or perfluoroalkyl (fluorocarbon) chemistries, hence the classification of siloxanes (silicon-oxygen bond).\\
The siloxane backbone is more flexible than a carbon backbone. This flexibility is attributed to the large $\mathrm{Si-O-Si}$ bond angle of $143^\circ - 150^\circ$ \cite{mark_conformations_1978, weinhold_nature_2011, chen_omniphobic_2023} which separates the pendant dimethyl groups. In addition, the alternating oxygen atoms do not allow side groups, resulting in low rotational energy around the backbone ($3.3~\mathrm{kJ/mol}$) - e.g., lower than what is noted between carbon-carbon bonds with methylene $\mathrm{CH_2}$ configurations ($13.8~\mathrm{kJ/mol}$) that typically occur in comparable organic polymers, such as polyethylene glycol \cite{colas2005silicones}. As a result, PDMS chains are highly flexible, and readily re-orientate to a minimal energy state, while possessing a low glass transition temperature ($\approx-150^\circ~\mathrm{C}$)\cite{volkov_polydimethylsiloxane_2015}.\\
The organic content in PDMS is relatively low compared to other silicones due to the small organic methyl group ($\mathrm{CH_3}$, is the smallest organic group). The apolarity of the methyl group yields weak interactions with polar solvents such as water and alcohols \cite{lee_solvent_2003, noll__1968}. This is reflected in the low solubility which, in the case of water, is only $\approx 770~\mathrm{ppm}$ \cite{tamai_molecular_1995,watson_behaviour_1996,harley_thermodynamic_2012}. Vice versa, PDMS is even insoluble in water \cite{amini_preventing_2017}. The poor interaction strength between PDMS and water, however, does not hamper the diffusive mobility of $\mathrm{H_2O}$ molecules in PDMS ($\approx 2\times10^{-9}~\mathrm{m^2/s}$)\cite{tamai_molecular_1994,harley_thermodynamic_2012} and is comparable with many other liquid-liquid diffusivities.\\
 The weak interaction strength with many solvents \cite{lee_compatibility_2004} results in low PDMS surface energy. Consequently, adhesion between contacting liquids and PDMS is relatively weak. Water repels from PDMS surfaces as reflected in the contact angles ranging between $90^\circ-110^\circ$ \cite{mata_characterization_2005,haubert_pdms_2006,kim_pdms_2011,qiu_contact_2014,seghir_extended_2015,ruben_oxygen_2017}. The flexibility of PDMS chains maintains slippage and thus provides mobility for contacting drops \cite{zhangnewby_effect_1997,henot_comparison_2017,yu_thermalinduced_2019,scarratt_how_2019}. Depending on the surface type, aqueous and organic drops can slide off when tilting the surface by a few degrees. However, on some PDMS surfaces drops remain pinned even when tilting the surface by $90^\circ$.\\
PDMS chains can be adjusted in length (\textit{i.e.}, molecular weight), cross-linkage, and chemical/physical surface attachments (“grafting”). Such variations shape distinctive surface types and features including viscosity, elasticity, film thickness, and durability. 

\subsection{Liquid-Infused Surfaces} 
The design of liquid-infused surfaces is to some extent nature-inspired: carnivorous plants such as the \textit{Nepenthes} pitcher plant utilize a combination of leaf texture and lubrication (the plant uses water as a lubricant) to trap insects such as ants \cite{bohn_insect_2004}. Liquid-infused surfaces comprise a micro- or nanostructured texture, infiltrated with liquid PDMS oil, also termed silicone oil. Typically, silicone oils are PDMS chains terminated with trimethyl groups, Fig \ref{fig:pdms_strucutre}a. The end groups are nonreactive; therefore PDMS remains liquid with a linear chain topology \cite{moretto_silicones_2000}, Fig. \ref{fig:pdms_strucutre}a, left. The oil viscosity is tunable \textit{via} the number of monomer units in the chain and ranges from $1~\mathrm{mPa~s}$ to $10~\mathrm{kPa~s}$ \cite{gelest_conventional_}. Short PDMS chains (“oligomers” \cite{jenkins_glossary_1996}) with $n<100$ monomer units are typically very mobile with low viscosities of $\eta\leq50~\mathrm{mPa~s}$. Due to the low surface tension of PDMS oils, almost all surfaces are “siliconephilic”, enabling rapid spreading of oil into the porous, micro-/nanostructure textures \cite{owen_why_1981,smith_droplet_2013,seiwert_coating_2011}. Micro- and/or nanostructures are added to the surface to enable stronger oil retention by capillary forces \cite{kim_hierarchical_2013}. These surfaces are summarized under the surface class named “SLIPS” (slippery liquid-infused porous surface) or “LIS” (lubricant-infused surface) \cite{wong_bioinspired_2011, lafuma_slippery_2011}. The oil provides lubrication, resulting in static friction forces orders of magnitude lower than dry friction. Most liquid drops slide off when tilting the surface by merely a few degrees. However, the infused oil can also be taken along, inducing degradation of the coating.\\ 
Various strategies for oil replenishment \cite{baumli_challenge_2021}, such as microfluidic setups \cite{baumli_flow_2019} or lubricant reservoirs \cite{wexler_sheardriven_2015,zhou_chemically_2024} have been implemented to enhance the lifetime of the surface.

\subsection{Elastomeric Surfaces} 
Crosslinked PDMS coated on substrates forms elastomeric surfaces, Fig. \ref{fig:pdms_based_coatings}b. Vinyl- or hydroxyl-terminated PDMS chains (Fig. \ref{fig:pdms_strucutre}a, right) are reactive and widely used in cross-linking reactions. In Sylgard 184 (a commercial PDMS elastomer), these reactive PDMS chains are mixed with a platinum-catalyzed methylhydrogensiloxane, $\mathrm{(HCH_3)SiO}$, cross-linker. In a hydrosilylation reaction, these shorter siloxane-based cross-linker chains ($n \approx 10$) undergo an addition reaction with the vinyl groups on the PDMS chain ends \cite{bogdan_hydrosilylation_}, Fig. \ref{fig:pdms_strucutre}b. Other derivatives of the synthesis exist. For example, the vinyl group can become incorporated in the siloxane backbone or other cross-linking agents may be utilized to realize addition and condensation crosslinking-reactions \cite{mani_crosslinking_2009,chasse_crosslink_2012,adam_static_1997,ciubotaru_silicones_2022}.\\ 
The mixture of activated PDMS chains and cross-linker is coated on a surface by e.g., spin-coating, dip-coating, or blade-coating \cite{xiong_spin_2015,butt_thinfilm_2022,yilgor_facile_2012}, before the reaction proceeds. Crosslinking reactions (independent of their formulation) will eventually arrive at the “gel point”. The gel point is achieved when a percolated bulk network forms and the material loses fluidity \cite{flory__1953}. Physically, this is an equilibrium state where the original PDMS liquid mixture becomes a soft solid (\textit{i.e.}, a gel). Three specific changes to physicochemical properties arise. i) Thermal rheology; as with other thermoset polymers, once a PDMS elastomer reaches the gel point, it can no longer reflow upon heating (unlike thermoplastics). This occurs above the glass transition temperature (or even melting point) of an equivalent linear (identical $\mathrm{M_w}$, uncrosslinked) PDMS chain. ii) Solvation; prior to gelation, the PDMS liquid mixture is readily soluble in certain solvents (e.g., hexane or toluene \cite{lee_solvent_2003}). However, post-gelation, cross-linked PDMS will experience swelling instead. The cross-links preserve the integrity of the elastomeric polymer network. iii) Mechanical; passing the gel point induces an abrupt increase of the viscosity and the emergence of elasticity \cite{chambon_stopping_1985}. Depending on the cross-linker to oligomer ratio, the material obtains a stiffness between $1~\mathrm{kPa}$ and $10~\mathrm{MPa}$ \cite{gutekunst_influence_2014,johnston_mechanical_2014,seghir_extended_2015,style_stiffening_2015,masterton_influence_2019}.\\ 
Even under optimal stoichiometric conditions, not all reactive chains will be crosslinked to the network \cite{lee_solvent_2003}. At least 3$\%$ of the chains will remain free within the network \cite{lee_solvent_2003}. These free chains alter the material properties. They can accumulate at the coating surface towards air and in the wetting ridge\cite{flapper_reversal_2023} where they separate from the network \cite{jensen_wetting_2015,cai_fluid_2021}. This has consequences for static and dynamic wetting \cite{hourlier-fargette_role_2017,wong_adaptive_2020,hauer_phase_2023}.

 \subsection{SOCAL Surfaces} 
The free chain ends of PDMS can be covalently anchored to a surface, creating a nanometer-thin coating \cite{teisala_grafting_2020,liu_onestep_2021,khatir_molecularly_2023,rostami_dynamic_2023,abbas_silicone_2023}, Fig. \ref{fig:pdms_based_coatings}c. Most chains are anchored at one end while the other remains free. Few chains can be grafted on both ends, forming chain loops \cite{litvinov_structure_2002,sakurai_preparation_2014}. The free-moving end is almost as flexible as liquid oligomers or oils and the surface interface is comparably slippery. Therefore, these surfaces are often called “liquid-like” or “SOCAL” (slippery, omniphobic, covalently attached, liquid) surfaces.\\
Surface anchoring is done either by a “grafting-from” or a “grafting-to” reaction. “Grafting-from” implies that chain polymerization is initiated at the anchor site on the surface and proceeds by adding PDMS monomer units. In a “grafting-to” reaction, an already polymerized chain binds to the surface \cite{krumpfer_rediscovering_2011,wang_covalently_2016,teisala_grafting_2020,liu_onestep_2021}. “Grafting-from” can be attained by e.g., highly reactive chloro-terminated siloxanes (monomers or oligomers). Long PDMS chains ($n > 1000$) will experience “grafting-to” anchoring. In both instances, they are reported to form covalent bonds and create molecularly thin layers \cite{milner_polymer_1991}.\\
Surface anchor sites are typically provided by hydroxyl groups, Fig. \ref{fig:pdms_strucutre}c. Anchor sites can be created using chemical treatment, e.g., oxygen plasma exposure or alkali activation \cite{bodas_hydrophilization_2007}. Many metals/metalloid oxides (e.g., $\mathrm{SiO_2}$, $\mathrm{TiO_2}$, $\mathrm{Al_2O_3}$, $\mathrm{NiO}$) naturally form hydroxyl groups under ambient water vapor exposure \cite{salmeron_water_2009}. Water dissociates on surfaces if acids (e.g. hydrofluoric acid) or alkali (e.g. sodium hydroxide) are used. Spontaneous dissociation of water on metal surfaces has also been reported \cite{ketteler_nature_2007,yamamoto_hydroxylinduced_2007}, alongside the spontaneous scission of covalent bonds in long PDMS chains \cite{teisala_grafting_2020}.\\
Grafted PDMS chains exist in a stretched or a collapsed state, depending on the affinity to the surrounding solvent and the chain density \cite{binder_polymer_2012}. Polar liquids (\textit{i.e.}, water) collapse chains as hydrophobic aggregation occurs to minimize surface energy. Organic liquids can stretch out chains if they are preferably solvated under similar chemical affinity. Anchored chains swell when exposed to PDMS oil. These solvent-dependent chain conformations lead to unique wetting behaviors: While both advancing $ \theta_a$ and receding angles $ \theta_r$ follow trends expected by surface tension variations (\textit{i.e.}, high for water, low for organic liquid), contact angle hysteresis $\Delta \theta = \theta_a- \theta_r$ is low for both liquids \cite{wooh_silicone_2016,barrio-zhang_contactangle_2020}. 
This is attributed to the nanometric smoothness of such surfaces and the flexibility exhibited by grafted chains. At the molecular level, grafted chains rotate almost freely, reducing contact friction \cite{krumpfer_contact_2010} Thus, contact angle hysteresis $\Delta \theta$ remains relatively low. Notably, $\Delta \theta$ depends on the grafting density and grafted chain length (\textit{i.e.}, the molecular weight) \cite{krumpfer_rediscovering_2011}. For example, $\Delta \theta$ was measured to be lower on SOCAL surfaces with $6~\mathrm{kDa}$ chains, compared to both shorter ($0.77~\mathrm{kDa}$) and longer ($117~\mathrm{kDa}$) chains \cite{teisala_grafting_2020}.

\section{Soft Wetting Mechanisms}

\subsection{Static Wetting Ridge} 

 All surface types share the formation of an annular wetting ridge around the three-phase contact line \cite{style_elastocapillarity_2017,chen_static_2018,bico_elastocapillarity_2018,andreotti_statics_2020}. The wetting ridge is defined as the region that rises above the original unperturbed level of the surface. On liquid-infused surfaces, the wetting ridge consists of PDMS oil only \cite{peppou-chapman_life_2020,baumli_challenge_2021,hardt_flow_2022}, whereas on elastomers, it may consist of a mixture of crosslinked and liquid PDMS, depending on the swelling ratio \cite{jensen_wetting_2015,cai_fluid_2021}. On SOCAL surfaces, the wetting ridge will encompass the local stretching of the grafted polymer chains \cite{leonforte_statics_2011, thiele_gradient_2020}.  As PDMS chains are organized differently on the liquid-infused, elastomer, and SOCAL surfaces, the material forces in the wetting ridge differ, as well. Thus, each surface type forms a characteristic wetting ridge with a specific geometry. The wetting ridge significantly contributes to the overall drop morphology and the wetting dynamics.

\subsubsection{Liquid}

In addition to the hydrophobic nature of PDMS oil, the impregnated oil acts as a lubricant. The film lubrication reduces the (static) friction of contacting objects (both solids and liquids) on the surface immensely. Low friction typically leads to excellent self-cleaning properties  \cite{lafuma_slippery_2011}. Furthermore, lubricant allows surfaces to “self-heal” from abrasion damages by the replenishment and re-flow of oil. However, once the PDMS oil is partially depleted from the surface, contacting drops are no longer lubricated and lose their mobility \cite{vega-sanchez_slightly_2022}. Maintaining the PDMS oil on the surface is, hence, the most important priority for the design of liquid-infused surfaces \cite{kim_hierarchical_2013,pham_droplets_2018,wong_capillary_2020}.\\
Oil depletion is already triggered by contacting water drops that induce the formation of annular wetting ridges, Fig. \ref{fig:wetting_ridge}a. Wetting ridges formed on lubricant-infused surfaces can be visible to the naked eye, Fig. \ref{fig:pdms_based_coatings}a. Still, the characteristic sizes are usually in the micro-domain (ca. $200~\mathrm{\mu m}$), and optical magnification helps to reveal more detail, Fig. \ref{fig:wetting_ridge}a right.\\
For drops smaller than the capillary length, gravity can be neglected and capillary interactions between the lubricant and the water drop solely govern the morphology of the wetting ridge. In contrast to wetting on inert rigid surfaces, a single angle, such as the Young angle \cite{young_iii_1805}, is insufficient for this wetting configuration \cite{marchand_contact_2012}. On soft surfaces, the surface material (the lubricant) adapts to the wetting drop as the interfacial tension of the drop pulls on the surface. The surface becomes curved in the vicinity of the wetting ridge. In equilibrium, the (liquid) wetting ridge is in a so-called “Neumann configuration”, which implies a balance of the three fluid interfacial tensions [$\vec{\gamma}_w + \vec{\gamma}_o +  \vec{\gamma}_{ow} = 0,$ subscripts correspond water-ambient, oil-ambient, and oil-water], Fig. \ref{fig:wetting_ridge}a left box. This tension balance yields a closed triangle, characterized by the threeassociated Neumann angles $\vartheta_1$, $\vartheta_2$, and $\vartheta_3$ \cite{princen__1969}. The angles are related to the interfacial tensions as 

\begin{equation}
    \label{eq:neumann_relation}
    \frac{\gamma_{w}}{\sin{\vartheta_1}} = \frac{\gamma_{o}}{\sin{\vartheta_2}} = \frac{\gamma_{ow}}{\sin{\vartheta_3}}.
\end{equation}

\noindent Drops of organic liquids (e.g., perflourinated oils) fulfill this condition \cite{jeon_moving_2023}. Due to the high surface tension of water, however, $\gamma_w > \gamma_o + \gamma_{ow}$, and hence no Neumann configuration exists., and we expect the oil to fully spread around the drop (cf. Sec. Free Chain Cloaking Layer). In other situations, the height of the wetting ridge (or the location of the three-phase contact line) keeps growing until the lubricant pressure in the wetting ridge matches the lubricant pressure in the solid textures. In equilibrium, each of the fluid interfaces assumes a minimal energy surface \cite{delaunay_surface_1841} where the mean curvature is constant \cite{gunjan_droplets_2020,lopez_complete_1989,meeks_classical_2011}. The oil interface deforms significantly in the surroundings of the wetting ridge but straightens, afar. The “catenoid” surface approximates all these properties. This is possible because the surface curvature in the two principal normal directions has equal magnitude but opposite signs, such that they perfectly cancel out. \\
While the classical Young equation/angle fails for lubricant-infused surfaces, a similar angle can be recovered in the limit when the wetting ridge is small compared to the size of the drop, Fig. \ref{fig:wetting_ridge}a center. This limit typically prevails in the “starved limit” when the infused PDMS oil is scarce. The resulting “pseudo Young angle” $\theta_{app}^S$ \cite{semprebon_apparent_2016} is

\begin{equation}
    \label{eq:starved_lis_ca}
    \cos{\theta_{app}^S}  = \frac{\Upsilon_{s}^{\mathrm{eff}} - \Upsilon_{sl}^{\mathrm{eff}}}{\gamma_w }.
\end{equation}

\noindent The surface tension of the solid texture and the PDMS oil are $\Upsilon_{s}$ and $\Upsilon_{sl}$, respectively. The solid texture takes up a fraction $\varphi$ of the overall surface. Thus, effective surface tension can be introduced, i.e., $\Upsilon_{s}^{\mathrm{eff}} = \Upsilon_{s} \varphi + \gamma_{o} (1-\varphi)$ for the solid texture and $\Upsilon_{sl}^{\mathrm{eff}} =  \Upsilon_{sl} \varphi + \gamma_{ow} (1-\varphi)$ for the surface-bound PDMS.\\
Moreover, in the starved limit, the shape of the wetting ridge is not only defined by capillarity. Disjoining pressures must also be considered when the oil film becomes very thin ($<\approx100~\mathrm{nm}$) because the oil-ambient interface starts to interact with the oil-solid interface \cite{israelachvili__1991}. Such interactions are characteristic of very thin films and are summarised in the “disjoining pressure” framework \cite{israelachvili__1991}. The interactions take place on a molecular level and can stem from a combination of van der Waals interactions, electrostatic attraction or repulsion, and steric forces \cite{butt_surface_2018}. For static PDMS oil films, the interactions are usually purely dispersive, and the disjoining pressure is 

\begin{equation}
    \Pi(h) = -\frac{A_H}{6\pi h^3},
\end{equation}

\noindent where $h$ is the film thickness and $A_H$ is the Hamaker constant. $A_H$ depends on the involved materials and can be approximated with the Lifshitz theory \cite{tabor_direct_1969}. In general, the disjoining pressure can be either attractive (unstable film) or repulsive (stable film). For PDMS oil on a solid-textured substrate $\mathcal{O}(A_H) \approx -10^{-19}~\mathrm{J}$ (cf. \cite{bergstrom_hamaker_1997} for a comprehensive material list). In case of severe film starvation, the disjoining pressure needs to be considered within the total free energy \cite{tress_shape_2017}. A consequence is that the oil interface assumes shapes, different from the catenoid \cite{gunjan_droplets_2020}.\\
When the surface holds sufficient PDMS oil, the contact angle defined in Eq. \eqref{eq:starved_lis_ca} breaks down because the assumption of a small wetting ridge no longer holds. The interface of the ridge now contributes to the total free energy. It becomes challenging to define a baseline from which to measure the (pseudo) angle, as the wetting ridge conceals the footprint of the drop. In practice, apparent contact angles $\theta_{app}$ are utilized, which are defined as the angle of the drop at the Neumann point to the horizontal plane, Fig. \ref{fig:wetting_ridge}a left. The size of the large ridge $\lambda_l$ and the size of the drop $r$ are related to the difference between $\theta_{app}$ and $\theta_{app}^S$ \cite{semprebon_apparent_2021} per

\begin{equation}
    \lambda_l = r \left( \cos{\theta_{app}} - \cos{\theta_{app}^S} \right).
\end{equation}

\subsubsection{Elastic}

\begin{figure*}
    \centering
    \includegraphics[width=1\textwidth]{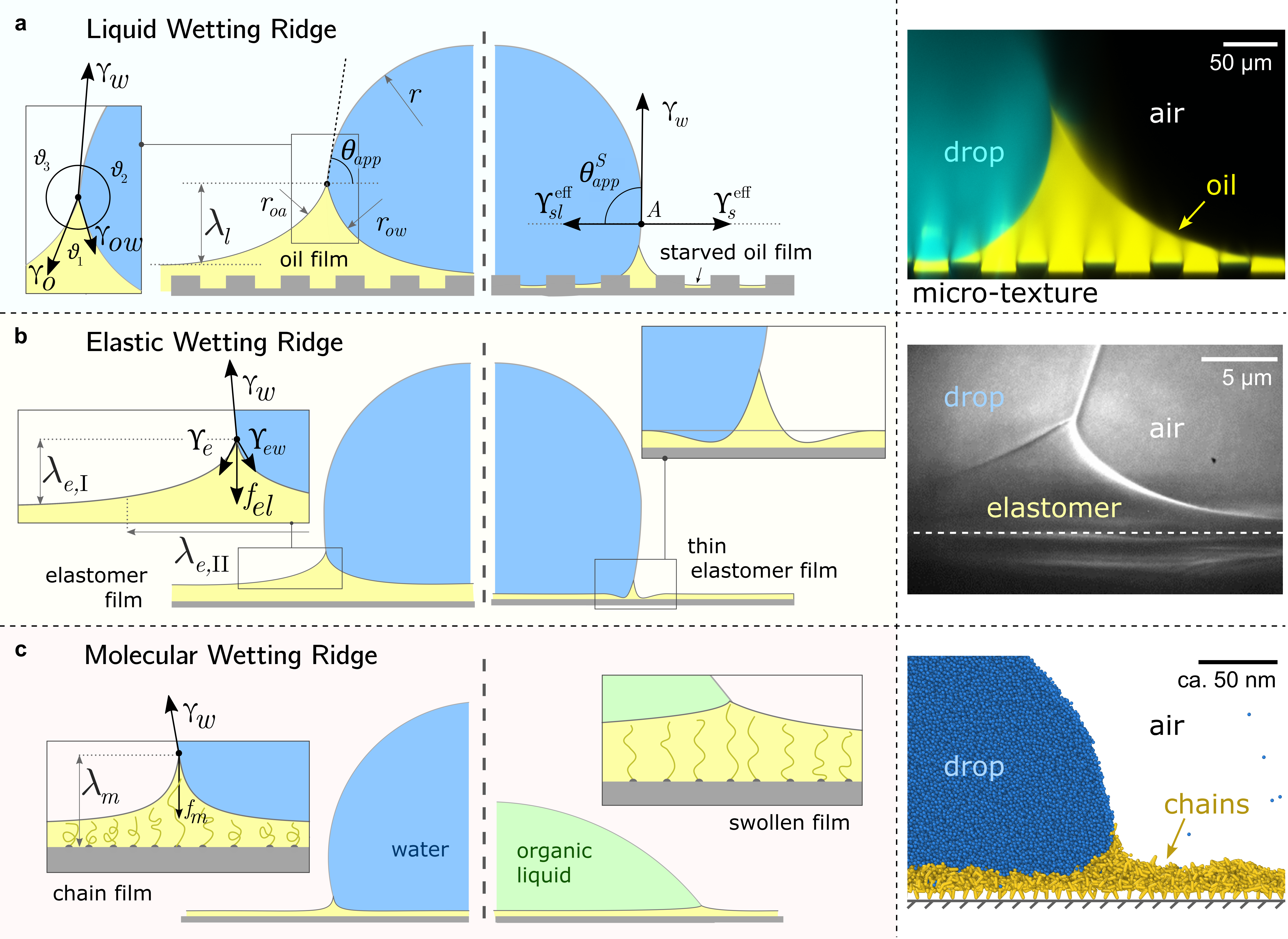}
    \caption{\label{fig:wetting_ridge}Different manifestations of the wetting ridge on different PDMS surfaces. a) Liquid wetting ridge on a liquid-infused surface. Neumann balance at the three-phase contact line (box) and the apparent contact angle $\theta_{app}$. The right half illustrates the starved limit and the pseudo-Young's balance. Macroscopically, the origin of the surface tension vectors (point $A$) aligns with the substrate baseline. The image on the right shows a liquid wetting ridge, taken with a fluorescence confocal microscope, adapted from \cite{baumli_challenge_2021}. b) Elastic wetting ridge on an elastomeric surface. Bulk elasticity creates a compliance force $f_{el}\propto G\lambda_{e,\mathrm{I}}$, resulting in wetting ridges smaller than the liquid ones. The right half shows material dimpling around the wetting ridge when the surface coating is very thin. The right image shows an elastic wetting ridge, taken with an x-ray microscope on elastomeric PDMS ($3~\mathrm{kPa}$), adapted from \cite{park_visualization_2014}. c) Molecular wetting ridge on a SOCAL surface. The entropic force of the PDMS chain $f_{m}$ limits the surface tension-induced chain stretch. Right half shows surface swelling when liquid-PDMS affinity/solubility is high. The right image shows a molecular wetting ridge, computed with coarse-grained molecular dynamics simulation, adapted from \cite{badr_cloaking_2022}.}
\end{figure*}

Wetting ridges also occur on elastomeric surfaces \cite{lester_contact_1961}. However, the mechanisms that govern the “elastic wetting ridge” are more complex than those governing liquid ridges. Capillary forces around the perimeter of a wetting drop induce mechanical stress in the PDMS material $(\mathbf{\sigma})$. Consequently, the material around the three-phase contact line deforms to a wetting ridge, similar to the liquid wetting ridge. In addition to the surface tension of the PDMS surface, the elasticity of the network bulk contributes to the ridge geometry. Depending on the bulk stiffness, wetting ridges on elastomers can grow up to tens of microns. With the steady improvement of optical techniques, elastic wetting ridges have been resolved with an increasing level of detail. Such modern state-of-the-art techniques are e.g., shadowgraphy, interferometry, and laser scanning confocal microscopy (cf. Methods and Techniques, \cite{carre_viscoelastic_1996,park_visualization_2014,gerber_wetting_2019,jerison_deformation_2011}, and Fig. \ref{fig:wetting_ridge}.) \\
The shape of the elastic wetting ridge can be found by balancing the bulk elasticity and the surface stress of the network to the imposed surface tension of the water drop at the three-phase contact line. The surface tension of the water pulls on the surface, acting like a point/line load on the PDMS material, reading

\begin{equation}
    \vec{\gamma}_w=\gamma_w \left(\cos{\theta}\vec{t} + \sin{\theta}\vec{n} \right) \delta\left\{x-x_{\mathrm{cl}}\right\}.
\end{equation}

\noindent Here, $\vec{n}$ and $\vec{t}$ denote the surface normal and tangential vectors respectively. $\delta\left\{x-x_{\mathrm{cl}}\right\}$ denotes the Dirac delta function and $x_\mathrm{cl}$ localizes the horizontal position of the three-phase contact line. The surface stress balance is then

\begin{equation}
    \label{eq:surface_stress_balance}
    \mathbf{\sigma}\vec{n} -\Upsilon_e \left(\nabla \cdot \vec{n} \right)\vec{n} = \vec{\gamma}_w.
\end{equation}

\noindent The spring-load system is a simple analogy to this wetting ridge where the resulting displacement - \textit{i.e.}, the height of the wetting ridge $\lambda_{e,\mathrm{I}}$ \cite{roman_elastocapillarity_2010, mora_capillarity_2010} - scales with the drop surface tension pulling on the surface per

\begin{equation}
    \label{eq:elastocapillary_lenght_I}
    \lambda_{e,\mathrm{I}} \propto \frac{\gamma_w}{G} \sin{\theta}.
\end{equation}

\noindent The compliance of the wetting ridge is characterized by $G^{-1}$, where $G$ is the shear modulus of PDMS. PDMS elastomers have a hydrophobic contact angle $\theta$ between $90^\circ-110^\circ$ \cite{mata_characterization_2005,haubert_pdms_2006,kim_pdms_2011,qiu_contact_2014,seghir_extended_2015,ruben_oxygen_2017}, Fig \ref{fig:pdms_based_coatings}b. Here, the solid surface tension $\Upsilon_e$ is considered symmetric around $x_\mathrm{cl}$. This, however, is not fulfilled as PDMS-air and PDMS-water interfaces have differing surface tension ($\Upsilon_{el}\approx 40~\mathrm{mN/m}$ and $\Upsilon_{e}\approx 20~\mathrm{mN/m}$). To match the asymmetric surface tension around the contact line, Eq. \eqref{eq:surface_stress_balance} can be adjusted with Heaviside functions. Note, the asymmetric surface tensions (dry vs. wetted) is not only invoked by water but is general for most liquids. \\
Solutions of this system of equations (and variations) show that the competition between the bulk elasticity $(G)$ and the capillary point load $(\gamma_w \sin{\theta})$ produces shapes that are comparable with experimental measurements \cite{park_visualization_2014}.\\ Considering only these two components, however, produces a singularity at the contact point of the load. The surface tension of the network material $\Upsilon_e$ regularizes this singularity by opposing the deformation with plastic stresses \cite{marchand_contact_2012,style_static_2012}. The length scale at which the surface tension-regularization emerges is  
 
 \begin{equation}
     \lambda_{e,\mathrm{II}} \propto \frac{\Upsilon_e}{G}.
 \end{equation}
 
\noindent When the wetting drop induces an elastic wetting ridge where $\mathcal{O}(\lambda_{e,\mathrm{I}})$ is below $\mathcal{O}(\lambda_{e,\mathrm{II}})$ the Neumann configuration [Eq. \eqref{eq:neumann_relation}] is recovered. The ratio of the two elastic length scales the strain $\left[\epsilon = \lambda_{e,\mathrm{I}}/\lambda_{e,\mathrm{II}}\right]$ and demarcates the two limits in which the stress can be treated:
 
 \begin{equation}
    \epsilon \propto \frac{\gamma_w}{\Upsilon_e}\sin{\theta} \begin{cases}
        \ll 1 &\mathrm{Neumann}\text{-}\mathrm{limit,}\\
        \gg 1 &\mathrm{bulk elasticity}\text{-}\mathrm{limit}.
    \end{cases}
\end{equation}

Straining solid (crystalline) surfaces can alter the molecular surface structure, yielding a local dependency of the surface energy on the strain \cite{lee_surface_2018}. Consequently, the surface energy and the surface tension are not equal [$\gamma_e\neq\Upsilon_e$]. The surface energy is a function of the surface strain $\gamma_e(\epsilon_\Sigma)$, and the relation to the surface tension is 

\begin{equation}
    \label{eq:shuttle_worth_eq}
    \Upsilon_e = \frac{\mathrm{d}}{\mathrm{d}\epsilon_\Sigma}\left[(1+\epsilon_\Sigma)\gamma_e(\epsilon_\Sigma)\right]. 
\end{equation}

\noindent Eq. \eqref{eq:shuttle_worth_eq} is known as the Shuttleworth equation \cite{shuttleworth_surface_1950,andreotti_soft_2016}. While originally introduced in 1950, it obtained recent momentum in the wetting community. The Shuttleworth effect was utilized to show anisotropic wetting behavior on unidirectionally stretched PDMS surfaces \cite{snoeijer_paradox_2018,vangorcum_spreading_2020,smith-mannschott_droplets_2021} and differing dependencies of $\Upsilon_e$ and $\gamma_e$ on the cross-link density \cite{zhao_role_2022}. Other studies found contradictory behavior and the absence of a $\gamma_e(\epsilon_\Sigma)$ dependency \cite{schulman_surface_2018}. Their observations were rationalized by only partially stretched elastomers (\textit{i.e.}, below the max. chain length) or oil-swollen elastomers where mobile chains at the surface enable the relaxation of strain-induced local molecular surface inhomogeneities \cite{rubinstein_elasticity_2002, liang_surface_2018,hourlier-fargette_role_2017}. In that sense, the PDMS elastomer surfaces resemble rather a liquid interface than a solid one. However, surface modifications, e.g. oxidation of surface molecules, can create a thin “surface skin” that is chemically and mechanically different from the bulk PDMS \cite{kim_longterm_2004, mills_mechanical_2008}. Upon stretching, the oxidized skin breaks, and non-oxidized PDMS molecules replenish the gap. Such surface-modified PDMS elastomers exhibit a strain-dependent surface energy/tension per Eq. \eqref{eq:shuttle_worth_eq}.\\
Further experimental findings indicate that the surface tension might not only depend on strain [per Eq. \eqref{eq:shuttle_worth_eq}] but on the strain rate, which could be interpreted as a form of complex surface rheology \cite{vangorcum_dynamic_2018}.

\subsubsection{Molecular \label{ch:mol_wetting_ridge}}

A wetting ridge on a SOCAL surface is tiny but possible: chains may stretch around the contact line, or ungrafted chains migrate freely to the contact line. Because of the nanometric thickness of the SOCAL surface, direct experimental observations of “molecular wetting ridges” have not been demonstrated. However, theoretical calculations - both on a molecular level and in a mean-field framework - predict their formation \cite{leonforte_statics_2011,thiele_gradient_2020,henkel_gradientdynamics_2021,badr_cloaking_2022}.\\
The molecular wetting ridge is, again, a result of the interplay of the capillary forces induced by the wetting drop and the material response of the surface. However, the motivation of the linear elasticity framework from a continuum mechanical point of view is less obvious as the length scales are much smaller. Since water is a poor solvent for PDMS \cite{tamai_molecular_1995,watson_behaviour_1996,harley_thermodynamic_2012}, the grafted chains are in a collapsed state, underneath the drop, Fig. \ref{fig:wetting_ridge}c left box. Deformations occur (at most) over the length of a chain (collapsed to stretched) in contrast to elastomers that undergo deformations far exceeding the lengthscales of the spacing between crosslinking nodes. Organic solvents are much better solvents\cite{lee_compatibility_2004} and enable chain stretching, leading to smaller wetting ridges, Fig. \ref{fig:wetting_ridge} center box. Further, the chain configuration introduces isotropy as the chains are not laterally reticulated. Therefore, the ridge deformation on SOCAL surfaces is usually considered in a thermodynamic framework of the individual polymer chains. The restoring force on the molecular chain level originates from entropy \cite{milner_theory_1988}. Stretched chains have a reduced conformational state. Thus, the entropy is low. This state is undesirable for the grafted chain. It strives to increase the possible conformations and maximize its entropy by recovering the unstretched state \cite{alexander_adsorption_1977,degennes_conformations_1980}. Notably, the thermodynamic consideration for a swollen chain recovers a linear relation between the deformation and the restoring force per chain

\begin{equation}
    f_m = k_B T \frac{H_B}{n b^2},
\end{equation}

\noindent where $k_B$ is the Boltzmann constant, $T$ is the temperature, $H_B$ is the thickness of the grafted chain layer, $n$ is the number of monomers in a chain, and $b$ is the size of a monomer. With this, the height of the molecular ridge is expected to scale as

\begin{equation}
    \lambda_{m} \propto  \frac{\gamma_w n b^2}{\sqrt{\rho} k_B T} \sin{\theta},
\end{equation}

\noindent where $\rho$ is the grafting density for the SOCAL surface $\left[\rho \approx 0.1~\text{chains/nm}^2\right]$ \cite{sakurai_preparation_2014}.

\subsection{Free Chains, Cloaking, and Swelling}

\subsubsection{Phase-Separated Wetting Ridge}

In Sylgard 184 PDMS elastomers synthesized under the standard recommended crosslinking procedure (10:1 base to crosslinker), around $5~\mathrm{\%wt}$ of the chains remain free. Free chains can be considered as a solvent that swells the crosslinked network \cite{regehr_biological_2009,lee_solvent_2003,glover_extracting_2020}. The volume ratio between the swollen network and the dry network defines the swelling ratio $Q=V_\mathrm{swollen}/V_\mathrm{dry}$. Depending on the degree of crosslinking, PDMS elastomers display different swelling ratios in equilibrium. For example, a gel prepared with Sylgard 184 with a 60:1 mixing ratio can be swollen up to $Q\approx 16$ with low molecular weight (oligomeric) PDMS chains \cite{cai_fluid_2021}.\\
Synersis - \textit{i.e.}, the extraction of free chains from the network onto the surface \cite{urata_selflubricating_2015,lavielle_lubrication_2021,cai_how_2022,park_superslippery_2023} - can cause lubrication \cite{hourlier-fargette_role_2017}. This essentially makes (swollen) PDMS elastomers similar to liquid-infused surfaces on which contaminants (such as water) hardly stick \cite{amini_preventing_2017}.\\
Synersis is typically induced by global effects such as temperature variations \cite{mizrahi_syneresis_2010} or the progression of the crosslink reaction \cite{lavielle_lubrication_2021}, affecting the entire surface of the PDMS elastomer. Wetting drops, however, trigger the accumulation of free chains locally around the three-phase contact line, caused by the surface tension of the drop \cite{hourlier-fargette_extraction_2018, wong_adaptive_2020}. On highly swollen PDMS elastomers, this can lead to the phase separation of free chains at the tip of the wetting ridge \cite{jensen_wetting_2015,cai_fluid_2021}. The extent of phase-separation (measured by the size of the separated tip, $h_\mathrm{oil}$) is proportional to the swelling ratio $Q$, \textit{i.e.}, the number of free chains in the gel. On PDMS-based SOCAL surfaces, such kinds of phase-separated tips have also been observed in molecular dynamics simulations \cite{badr_cloaking_2022}. Due to the small length scales of the surface, direct experimental visualization is still lacking.

\subsubsection{Cloaking Layer}

The surface tension of a wetting drop yields the formation of a wetting ridge and the accumulation of free chains (PDMS oil) \cite{jensen_wetting_2015,cai_fluid_2021}. A sharp three-phase contact line at the tip of the ridge implies a (Neumann) balance [Eq. \eqref{eq:neumann_relation}] of the competing surface tensions of the water ($\gamma_w$), the oil ($\gamma_o$), and water/oil ($\gamma_{ow}$). The Neumann balance is fulfilled with a negative “spreading coefficient”:

\begin{equation}
    S = \gamma_{w} - \gamma_{ow} - \gamma_o \leq 0.
\end{equation}

\begin{figure}
\centering
    \includegraphics[width=0.5\textwidth]{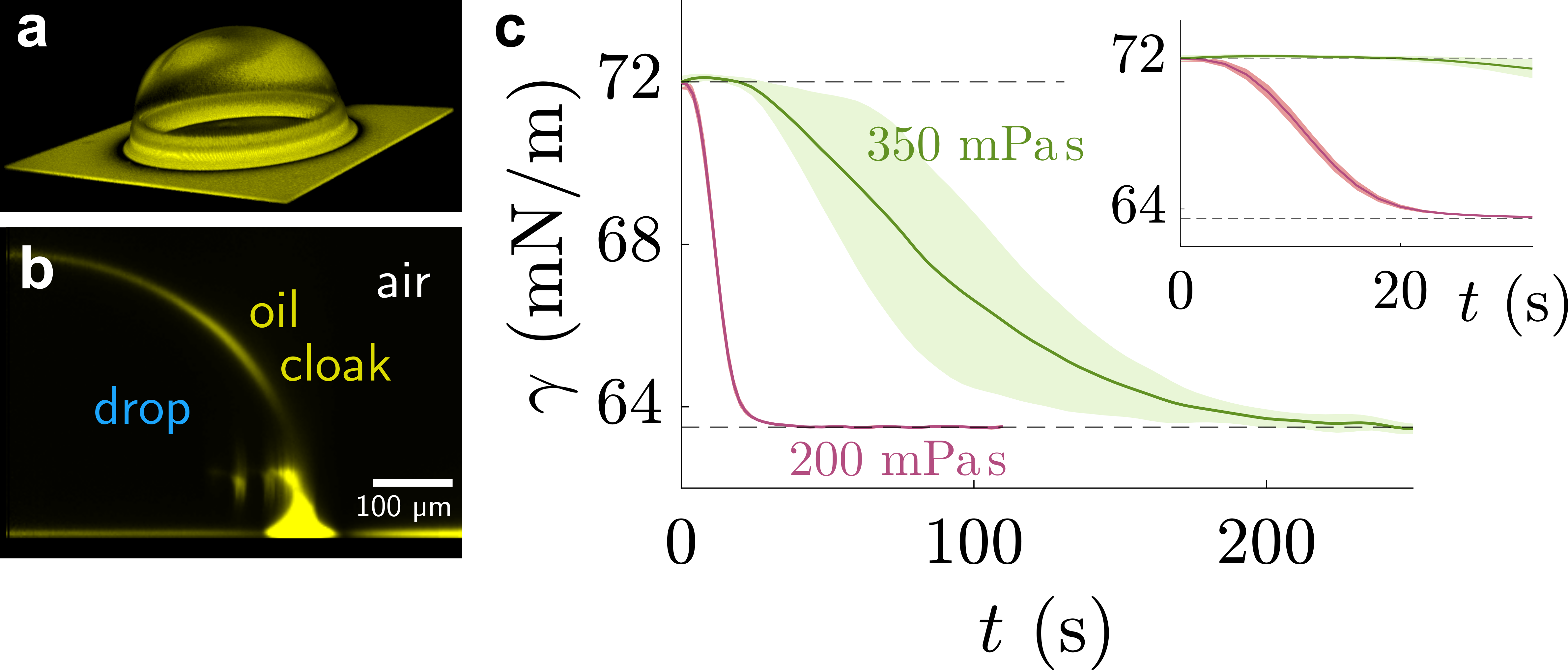}
    \caption{\label{fig:oil_cloak_lscm} Sessile drop on lubricated surface, cloaked by PDMS oil. \textbf{a)} Three-dimensional view of PDMS oil on the drop, obtained by laser scanning confocal microscopy. \textbf{b)} Side view of the angular averaged 3D stack. Images adapted from \cite{badr_cloaking_2022} c) Temporal evolution of pendant drop on PDMS elastomer with swollen with $200~\mathrm{mPa~s}$ (red) and $350~\mathrm{mPa~s}$ (green) PDMS oil. The shaded region shows the standard deviation of experimental repeats. Data adapted from \cite{naga_capillary_2021}.}
\end{figure}

\noindent Generally, $S\leq0$ indicates the presence of a contact line (partial wetting), and $S>0$ expresses the absence of a contact line in equilibrium (complete wetting). As all three interfaces are liquid, their surface tensions are easily accessible with direct measurements (cf. Shadowgraphy in Methods and Techniques) and for water drops, values are listed in Tab. \ref{tab:surface_tension_oil_water_air}. Notably, for these values, the spreading coefficient is positive $(7-16~\mathrm{mN/m})$, and thus, PDMS oil tends to spread on top of the water drop \cite{bergeron_monolayer_1996}. The PDMS engulfs the drop with a thin layer, Fig. \ref{fig:oil_cloak_lscm}a,b. This behavior is called “cloaking”, and significantly changes the wetting configuration \cite{sharma_enhanced_2022}. Due to the cloak, the surface energy of the drop cap changes, affecting the total free energy and consequently the wetting affinity of the drop and the contact angle, respectively. A cloaking layer of PDMS oil makes the Neumann configuration impossible. Instead, the effective surface tension of the drop changes.\\

\begin{table}[h!]
    \centering
    \caption{\label{tab:surface_tension_oil_water_air}Interfacial tension between water-air, PDMS oil-air, and PDMS oil-water interface at room temperature. Differences in the values can stem from differing PDMS chain lengths ($\eta \approx 5-300~\mathrm{mPa~s}$) and measurement errors.}
    \begin{tabular}{lc|c}
        Interface &  & Surface Tension $[\mathrm{mN/m}]$\\
        \hline
        water-air & $\gamma_w$ & $72$ \cite{hauer_wetting_2023}\\
        PDMS oil-water & $\gamma_{ow}$ & $40$ \cite{barca_silicone_2014}, $38$ \cite{teisala_wetting_2018}\\
        PDMS oil-air & $\gamma_{o}$ & $18-22$ \cite{teisala_wetting_2018}, $21$ \cite{barca_silicone_2014}, $25$ \cite{toor_reconfigurable_2018}\\
    \end{tabular}
\end{table}

Naively, the “cloaked” surface tension $\gamma_{c}$ of the drop effectively sums up from two interfaces (water-oil and oil-air) that replaces the uncloaked surface (water-air), \textit{i.e.},  $\gamma_{c} = \gamma_{ow} + \gamma_{o}$. However, the concepts of surface tension and surface energy consider shared surfaces of two bulk phases. Surface tension and energy result from excess energy that decays at some length scale towards the bulk phases \cite{butt_physics_2013}. The length (or the vicinity) in which the excess energy decays is also reflected in the stress isotropy, being isotropic in the bulk phases, and anisotropic near the interface \cite{nagata_molecular_2016}. Following this reasoning, $\gamma_c$ is only the sum of $\gamma_{ow}$ and $\gamma_{o}$  when the cloaking layer is thick enough and the surface tension has “bulk properties”. However, the layer can be substantially thinner, especially upon cloak formation \cite{kreder_film_2018}. In this case, the two interfaces (water-oil and oil-ambient) of the thin cloak layer start to interact with each other \textit{via} the disjoining pressure $\Pi(h_c)$ yielding stabilizing stress contributions which depend on the thickness of the cloak $h_c$. The stabilizing disjoining pressure contribution adds to the surface tension of the cloak with \cite{platikanov_thin_2005,hadjiiski_gentle_2002,gunjan_cloaked_2021}

\begin{equation}
    \label{eq:cloak_tensio}
    \gamma_c\left(h_c \right) = \gamma_{ow} + \gamma_{o} + \Pi(h_c)h_c  + \int_{h_c}^{\infty} \Pi(h) dh.  
\end{equation}

\noindent Note, that for $h_c \to\infty$, the disjoining pressure vanishes and $\gamma_{c} = \gamma_{ow} + \gamma_{o}$ is recovered. In the opposite limit, when $h_c\to 0$, $\gamma_c\to \gamma_w$ such that the surface tension of the uncloaked drop is recovered.\\
The formation of the cloak is a transient process, associated with time scales of the oils rooting in viscous or diffusive oil transport from the surface to the drop cap. Hence, the temporal evolution of the cloak formation can be measured: experimentally, this has been illustrated with pendant drop measurements, where drops hang on swollen PDMS elastomer surfaces \cite{hourlier-fargette_extraction_2018, nath_how_2020, naga_capillary_2021}. In these experiments, the surface tension of the hanging drop ($38~\mathrm{\mu l}$) changed from the characteristic $\gamma_{wa} = 72~\mathrm{mN/m}$ to  $\gamma_c\approx 64~\mathrm{mN/m}$ after more than $\approx 3$ minutes, Fig. \ref{fig:oil_cloak_lscm}c. Besides oil parameters (such as viscosity), the drop size influences the cloak formation time \cite{hourlier-fargette_extraction_2018}. 
The coupling in Eq. \eqref{eq:cloak_tensio} between $\gamma_c$ and $h_c$ indicates that one is a proxy for the other. Hence, the temporal evolution of $\gamma_c$ may be linked to the gradually increasing cloak thickness. Note, that $\gamma_{wo} + \gamma_o \approx 64~\mathrm{mN/m}$, Tab. \ref{tab:surface_tension_oil_water_air}, indicating that the cloak reached a bulk-like thickness for $t \to t_\infty \approx 3~\mathrm{min}$.

\subsubsection{Swelling}

Water and other polar liquids typically have a poor affinity with PDMS (cf. Structure and Properties). Other liquids (in particular organic ones) however, interact much stronger with PDMS. When such organic liquids wet a PDMS surface (LIS, elastic, or SOCAL), the wetting drop can "swell" the underlying PDMS substrate. The extent of swelling depends on the affinity between the wetting liquid and the PDMS. For some liquids (e.g. toluene or hexane) the liquid fully immerses into the PDMS, increasing the volume by more than 220 \% \cite{lee_compatibility_2004}. Swelling alters the material properties of the PDMS surface, e.g. the surface tension \cite{binks_mixing_1998,kleingartner_exploring_2013,butt_adaptive_2018}, the shear modulus \cite{li_liquidinduced_2021}, the viscosity, etc. Hence, swelling directly influences the static and dynamic wetting behavior, i.e., the contact angle $\theta$, the hysteresis $\Delta \theta$, and the dissipation (cf. Dynamic Wetting Ridge).\\
When the wetting drop moves, the swelling kinetics can compete with the sliding of the drop. This leads to a steady state where the degree of swelling is not equilibrated but depends on the sliding speed of the drop. In this state, the surface underneath the front and rear of the drop swell to different degrees, and the PDMS wettability exhibits local gradients. This induces a contact angle hysteresis that does not depend on the surface inhomogeneity but the swelling kinetics, the drop size, and the sliding speed \cite{butt_adaptive_2018}.

\subsection{Dynamic Wetting Ridge}

\subsubsection{Liquid}

\begin{figure}
    \centering
    \includegraphics[width=.5\textwidth]{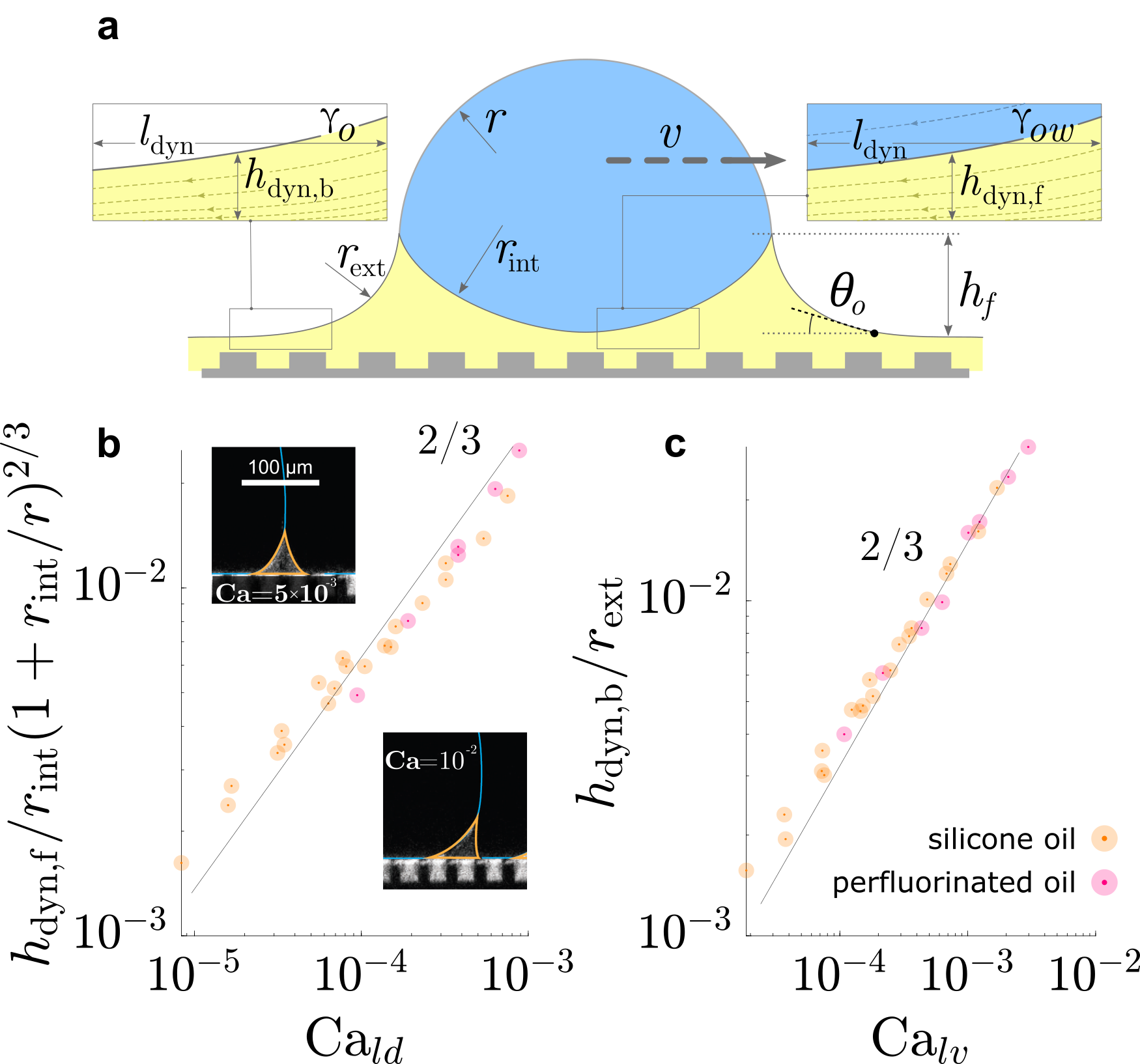}
    \caption{\label{fig:wetting_ridge_dyn} Dynamic geometries of liquid wetting ridge during drop sliding. The oil film underneath the drop and behind the drop, both, resemble a Landau Levich Derjaguin film. The dynamic film thickness a) underneath the drop front $h_\mathrm{dyn,f}$, and b) behind the drop rear end $h_\mathrm{dyn,b}$, both display the scaling $\propto\mathrm{Ca}^{2/3}$, characteristic for Landau Levich Derjaguin films. Data in b) and c) adapted from \cite{kreder_film_2018} and insets in b) adapted from \cite{keiser_drop_2017}.}
\end{figure}

Sliding drops experience friction, stemming from shear dissipation. On rigid surfaces, the source of dissipation is either found at the three-phase contact line (for small speeds) or at the footprint of the drop where a hydrodynamic boundary layer forms (at higher speeds). On PDMS surfaces, the main source of dissipation, however, can be localized in the PDMS layer instead of the drop. This is the case when the lubricant is more viscous than the drop \cite{keiser_drop_2017}. The wetting ridge at the three-phase contact line moves with the drop and the accumulated material in the ridge is constantly reorganized, Fig. \ref{fig:wetting_ridge_dyn}a. During reorganization, the material undergoes shear and strain that dissipates energy due to the viscous/viscoelastic nature of the PDMS material. Note, that shear is not only evoked in the ridge but on all parts of the PDMS surface that are in contact with the drop (\textit{i.e.}, the drop footprint); however, the dissipation in the wetting ridge is the dominant source of sliding friction \cite{daniel_oleoplaning_2017,keiser_drop_2017,keiser_universality_2020,roche_complexity_2024}. Hence, to quantify the sliding friction, it is sufficient to only focus on the wetting ridge.\\
In general, the dissipation $P_{\mathrm{diss}} = \int_{V_\mathrm{ridge}} \mathbf{\sigma}:\dot{\mathbf{\epsilon}}dV$, depends on the volume of the wetting ridge $V_\mathrm{ridge}$, the strain rate $\dot{\mathbf{\epsilon}}$ (\textit{i.e.}, the rate of reorganization), and the material stress $\mathbf{\sigma}$. The stress in the wetting ridge follows from material laws, and thus, changes with the fabric of the different surface types. PDMS oils with $\eta<20~\mathrm{Pa~s}$ are Newtonian fluids that show a linear relationship between stress and shear rate, $\mathbf{\sigma}=\eta\dot{\mathbf{\epsilon}}$. If the oil is very viscous it can exhibit shear-thinning behavior \cite{hadjistamov_determination_2008}, however, the drop mobility is significantly hampered in the Newtonian low-speed regime\cite{keiser_drop_2017,keiser_universality_2020}. Still, lubrication enables higher drop mobility on lubricated compared to non-lubricated surfaces which tend to be rougher and have a much larger number of surface defects that slow down the drop \cite{solomon_drag_2014}. The shear stress in the liquid wetting ridge scales as $\sigma \propto \eta v/h_r$. The dissipation rate is found by integrating the viscous stress over the ridge area. Furthermore, we must account for the variation of dynamic contact angle of the wetting ridge with speed, Fig \ref{fig:wetting_ridge_dyn}b,c), which leads to, $P_\mathrm{diss} \propto v \beta \phi r \left(\eta v/\theta_o \right)$. $\phi$ accounts for the texture density which is important as only texture contacts along the perimeter add to the overall friction. On parts where the ridge contacts oil, it is assumed that no significant shear builds up and can, therefore, be neglected. $\beta=\ln{(h_r/\alpha)}$ and $\alpha\approx 100~\mathrm{nm}$ account for a cut-off length \cite{huh_hydrodynamic_1971}. $\theta_o$ is the opening angle, formed between the horizontal plane and the wetting ridge. This angle changes with the speed according to Tanner's law $\theta_o \approx \left( \beta \eta v / \gamma \right)^{1/3}$ \cite{tanner_spreading_1979}. Including Tanner's law into the expression for the dissipative power yields $P_\mathrm{diss} \propto v \phi r \left(\beta  \eta v/\gamma \right)^{2/3}$. The ratio between the viscous stress and the surface force in the liquid wetting ridge are condensed in the capillary number $\Ca = \eta v/\gamma$. Hence, the dissipative scaling can be written compactly as 

\begin{equation}
    P_\mathrm{diss}^\mathrm{vis,I} \propto \phi v r \left(\beta  \Ca \right)^{2/3}.
\end{equation}

\begin{figure*}
    \centering
    \includegraphics[width=.95\textwidth]{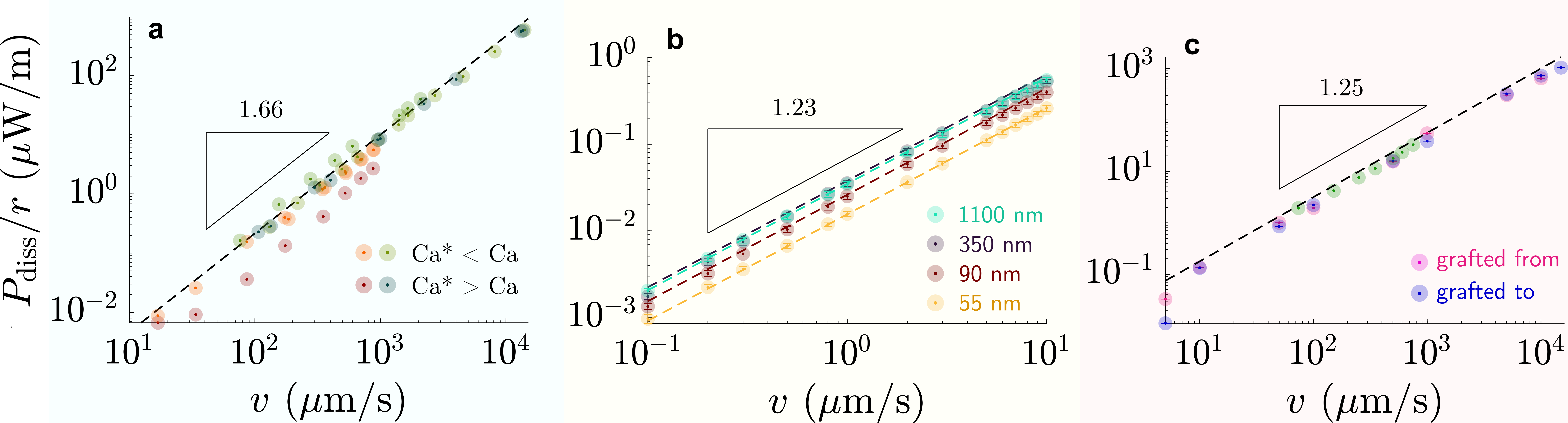}
    \caption{\label{fig:wetting_ridge_dissipation} Dissipation-speed profile ($P_\mathrm{diss}/r\propto v^k$) on a) liquid-infused silicone oil (green) $\eta=100~\mathrm{mPa~s}$ and perfluorinated oil (red) $\eta=30~\mathrm{mPa~s}$ ($k\approx1.66$), b) $G=250~\mathrm{kPa}$ elastomer ($k\approx1.23$), and c) $6~\mathrm{kDa}$ chain SOCAL surfaces ($k\approx1.25$). a) Bright colors correspond to cases where $\Ca^*<\Ca$ and dark colors to cases where $\Ca^*>\Ca$, with $\Ca^* = (h_\mathrm{p}/ r)^{3/2}$.b) The colors correspond to various coating thicknesses: green - $1100~\mathrm{nm}$, blackberry - $350~\mathrm{nm}$, red - $90~\mathrm{nm}$, yellow - $55~\mathrm{nm}$. c) Data points were measured on SOCAL surface, with chains: pink - grafted from, blue - grafted to substrate (c.f. PDMS based surfaces - structure and properties). Data reproduced from a) red \cite{daniel_oleoplaning_2017} green \cite{keiser_drop_2017,keiser_universality_2020}, b) \cite{khattak_direct_2022}, and c) green \cite{pilat_dynamic_2012} rest \cite{badr_dynamics_2024}.}
\end{figure*}

\noindent Despite its simplicity, this relation has been remarkably successful at capturing the scaling between dissipation and capillary number in experiments performed with various surface geometries and lubricant viscosities by independent groups, Fig. \ref{fig:wetting_ridge_dissipation}a.\\
On flat lubricated surfaces, the ridge meniscus behaves analogously to a Landau Levich Derjaguin film: the viscous stress elevates the film in the vicinity $l_\mathrm{dyn}$ behind the advancing and the receding wetting ridge \cite{degennes_capillarity_2010, daniel_oleoplaning_2017}. The elevated height scales as $h_\mathrm{dyn} \propto r\Ca^{2/3}$, and the vicinity length as $ l_\mathrm{dyn} \propto r \Ca^{1/3}.$ Depending on whether considering the advancing or receding ridge, $\Ca$ is formed with $\gamma_{ow}$ and $\gamma_o$, respectively. As many oil surfaces utilize a porous structure to maintain better retention of the oil film, $h_\mathrm{dyn}$ only forms when this height exceeds the height of the porous structure $h_\mathrm{p}$ \cite{seiwert_coating_2011}. Combining the dynamic $h_\mathrm{dyn}$ scaling with this inequality, produces the critical capillary number $\Ca^* = (h_\mathrm{p}/ r)^{3/2}$ that needs to be exceeded to form $h_\mathrm{dyn}$ \cite{keiser_universality_2020}. The associated (secondary) dissipation rate within this domain scales (non-linearly in $v$, Fig. \ref{fig:wetting_ridge_dissipation}a) as 

\begin{equation}
    P_\mathrm{diss}^\mathrm{vis,II} \propto \gamma v r \Ca^{2/3}.
\end{equation}

\noindent Since the scaling between power and the capillary number is the same for both dissipation in the Landau Levich Derjaguin film and at the foot of the wetting ridge, most experiments have the same scaling regardless of the dissipation mechanism. While this makes the scaling law remarkably universal, it also makes it difficult to directly deduce which of the two mechanisms is dominant.\\
At high capillary numbers ($\Ca>\approx 10^{-2}$), a different scaling law was observed between friction on velocity, with a lower exponent on the velocity. In this regime, drops begin to move more rapidly. The mechanism for this regime is still unknown. A tentative explanation that has been proposed is that the wetting ridge shape does not have time to fully develop in this regime \cite{li_rapid_2023}.

\subsubsection{Elastic}

On elastomeric surfaces, the wetting ridge dissipation is more complex \cite{shanahan_viscoelastic_1995,karpitschka_droplets_2015,oleron_morphology_2024,roche_complexity_2024}. Dissipation in elastomers is non-Newtonian and depends on the (oscillatory) strain rate $\omega$. The crosslinks add (visco)elasticity to the material fabric. For cross-linked PDMS in particular, the resulting material stress can be well described with the power law model proposed by Chasset and Thirion \cite{chasset__1965}, \textit{i.e.}, $\sigma \propto \epsilon G \left(\tau \omega \right)^m$. The moving wetting ridge moves with $\omega\propto v/\lambda_\mathrm{e,I}$\cite{karpitschka_droplets_2015}. $\tau$ is the rheological relaxation time and $m<1$ yields characteristic shear-thinning of PDMS \cite{ghannam_rheological_1998,petet_rheological_2021}. These two values vary depending on the degree of crosslinking and the number of free chains in the network \cite{batra_stress_2005}. Softer elastomers with more free chains usually have both smaller $\tau$ and smaller $m$. The strain in elastic ridges scales with the ratio of the two elastocapillary lengths $\epsilon\propto \lambda_\mathrm{e,I}/\lambda_\mathrm{e,II}$ and the strain rate with $\dot{\epsilon}\propto\epsilon\omega$. Together with the volumetric scaling $V \propto \lambda_\mathrm{e,I}^2$, the dissipative work in the viscoelastic wetting ridge scales as 

\begin{equation} 
    P_\mathrm{diss}^\mathrm{vel} \propto \gamma_w v r \epsilon^2 \Ca_e^m,
\end{equation}

\noindent Fig. \ref{fig:wetting_ridge_dissipation}b. Here, we introduce the “viscoelastic” capillary number $\Ca_e = \tau v/\lambda_{e,I}$. The thickness of the surface coating affects the shape of the wetting ridge when the length scales of both become comparable \cite{pericet-camara_solidsupported_2009, zhao_geometrical_2018}. Surface material that accumulates in the elastic wetting ridge cannot be replenished. Thus, the surface dimples around the wetting ridge, in lack of surface material, Fig. \ref{fig:wetting_ridge}b. Very thin coating surfaces also limit the size of the wetting ridge as the material is confined (or attached) to an underlying surface. The ridge geometry is in addition to $\lambda_{e,\mathrm{I}}$ and $\lambda_{e,\mathrm{II}}$ governed by the thin film height. The friction that builds up in the wetting ridge is consequently reduced, as the total dissipation volume, $V_\mathrm{ridge}$ decreases \cite{khattak_direct_2022}.\\
Wetting ridge relaxations that are different from expected viscoelastic materials were observed on PDMS elastomers having free oligomer chains \cite{xu_viscoelastic_2020}. The reorganization of the free chains inside the elastomer follows poroelastic behavior, which is typically slower than viscoelastic dissipation \cite{zhao_growth_2018}. Wetting ridges that move slowly dissipate energy in a different mode, as the material fabric can change significantly inside the ridge. As discussed earlier, a phase of pure liquid can form at the tip of the ridge where the dissipative picture resembles the liquid-infused case. The amount of phase separation scales inversely with the moving speed of the ridge \cite{hauer_phase_2023}. Thus, a complex back coupling between speed and dissipation is expected. While concise models for dissipation on swollen PDMS elastomers are missing, first clues suggest that the separation lowers dissipation: the accumulated oligomeric free chains have low viscosity and lubricate the drop and effectively reduce contact friction \cite{hourlier-fargette_role_2017, wong_adaptive_2020}. 

%\begin{figure}
%    \centering
%    \includegraphics[width=.3\textwidth]{Figures/Figure_elastomer_separation.png}
%    \caption{\label{fig:wetting_ridge_dyn_separation}\note[LH]{a) is from Xu PRL 2020 which shows ridge relaxation of swollen PDMS with interferometry. The deviation from power law indicated deviation from viscoelasticity.}}
%\end{figure}

\subsubsection{Molecular}

In SOCALs, a mobile drop sliding over the surface reorganizes the stretched chains in the ridge constantly. While low, the contact angle hysteresis is not zero. This indicates a non-symmetric process: the differing capillary force at the advancing and receding side leads to a non-zero net force as the drop moves. The force that builds up during sliding was measured with a drop friction force instrument \cite{badr_dynamics_2024}. This force scales with the drop speed approximately with $P_\mathrm{diss}\propto v^{5/4}$, Fig \ref{fig:wetting_ridge_dissipation}c. These measurements have some important implications: as the variation in synthesis does not appear to influence the effective friction force, therefore, the grafting density and chain length may not play a significant role. However, rigorous investigations are still lacking and decisive conclusions should not be drawn at this stage.

\section{Coating Comparison}

In this review, we discussed wetting behaviors on PDMS-based surfaces. We discuss how this relatively simple polymer can create a wide design space for surfaces, \textit{via} polymer-chain configurations, polymerization, and surface anchoring. The complexity in surface manufacturing varies for different surface variants: Creating liquid-infused surfaces without a micro-/nanometric texture is straightforward. It can be achieved by simply spreading PDMS oil on a surface. The process is equally facile for PDMS elastomeric surfaces, but only if no special requirements regarding the coating thickness exist. The “grafting-from” technique for SOCAL surfaces requires a more elaborate process such as “chemical-vapor-deposition” (CVD), requiring controlled atmosphere conditions, and sometimes toxic chemicals (e.g. chlorosilanes or alkoxysilanes).\\
The various surface designs open up a wide variety of concepts and mechanisms. Today, they are increasingly utilized and tuned for a plethora of applications. While similar at first glance, wetting mechanisms such as the wetting ridge develop differently on each PDMS-based surface variant. In the context of a specific application, this brings advantages and disadvantages. For example, self-cleaning is increasingly essential for preserving the energy efficiency of solar panels,
bio-contamination reduction of medical equipment, or repelling stains and dirt on outdoor equipment and textiles. For this, all three PDMS-based surface variants are conceptually suitable as they strive to provide drop mobility (low static friction), a feature that is essential for self-cleaning.\\
Liquid-infused surfaces have freely flowing PDMS oils providing excellent lubrication properties and high drop mobility. An important avenue for future research is to understand how particulate contaminants adhere to the lubricant layer and whether they enter the solid crevices. However, one limitation of liquid-infused surfaces is that lubricant is depleted by the (comparatively) large wetting ridge when they are exposed to harsh environments. Thus, to make these surfaces viable for technological applications, further work is needed to systematically understand the mechanisms of lubricant depletion. Elastomer surfaces, on the other hand, have good material retention and hence mechanical durability, provided by the crosslink network. This advantage, however, brings along higher dissipation margins and declined drop mobility. SOCAL surfaces seem to overcome both challenges, as retention is high due to covalent surface anchors, while dissipation remains low, although not as low as liquid-infused surfaces. This is attributed to a minuscule ridge and the inherent flexibility of molecularly anchored chains that behave in a liquid-like manner. Particulate matter hardly sticks and can be removed easily. However, once the surface experiences critical damage, the nano-thickness of the SOCAL surface can be permanently abraded away. Bulky liquid-infused and elastomer surfaces provide plenty of surface material that - in case of surface damage - can self-heal and increase the surface lifetime.\\

\begin{figure*}
    \centering
    \includegraphics[width=1\textwidth]{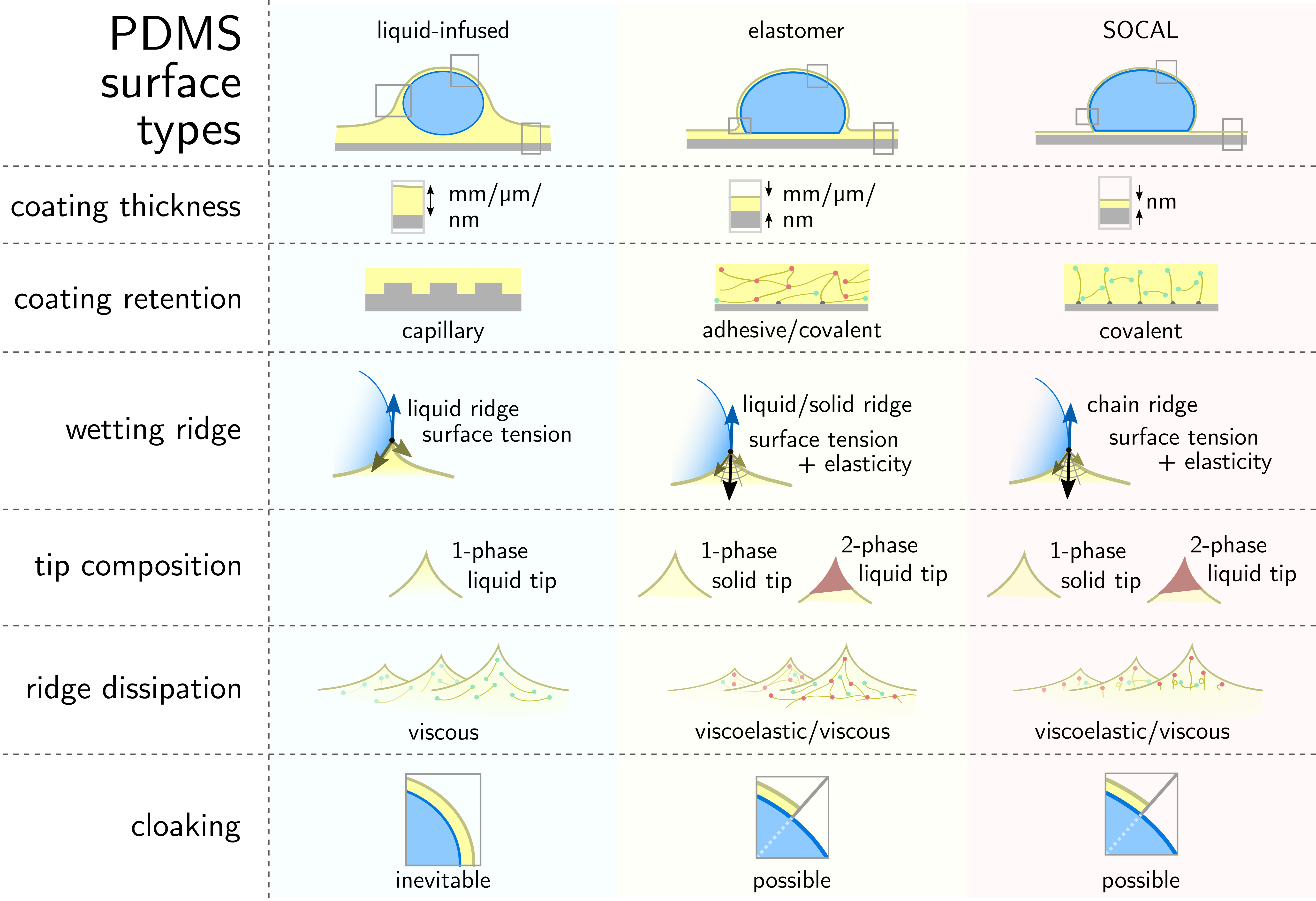}
    \caption{Comparison of PDMS-based surface structures and wetting mechanisms on the three variants. Illustrations adapted from \cite{hauer_wetting_2023}.}
\end{figure*}

\section{Open Questions and Perspectives}

PDMS surfaces share several similarities regardless of whether they are in the form of elastomers, SOCALS, or LIS. The wetting ridge is a pivotal element in static and dynamic wetting on each PDMS-based surface type. This is remarkable, considering that the length scale of the wetting ridge can vary by several orders of magnitude and the (macro)molecular configurations can differ significantly between the different variants. It remains an open question whether a unifying model can be developed to describe the wetting dynamics on these different surfaces. To make progress on that front, future studies need to investigate the reorganization of PDMS (oily, crosslinked, or surface-grafted) and the associated dissipation in the dynamic wetting ridge in detail.\\
The cloaking phenomenon has been observed experimentally on LIS and elastomeric surfaces. Yet, models for the dynamics of cloak formation remain lacking and it remains unknown whether the cloak influences the dissipation mechanisms. Even in equilibrium, the cloak bears an intriguing paradox for water drops:  While the spreading coefficient suggests the formation of a stable cloak, dispersive van der Waals (vdW) interactions counterintuitively indicate an attractive disjoining pressure, leading to an unstable film. This paradox may be due to other stabilizing long-range forces (e.g. electrostatic). Further work is needed to understand under which conditions the cloak is stable and whether it influences the dissipation mechanisms. A few considerations and open questions are suggested here for the three types of surfaces.

\begin{itemize}
    \item \textit{LIS} - The PDMS oil in LIS is usually a homogenous, Newtonian, liquid, suggesting a simple design space of the surface. Yet, several methods exist to create the underlying solid substrate and to imbibe the lubricant, leading to a large zoo of LIS designs. Friction laws have been proposed for drops moving on LIS. Still, direct (experimental) validation of the dissipation mechanism in the wetting ridge remains lacking and it is still unclear whether and how the exact geometry of the solid substrate influences the dissipation mechanism. Depletion of lubricant over time remains the biggest challenge that needs to be tackled to design long-lasting PDMS surfaces. The wetting ridge and the cloak are significant causes of depletion when drops slide off lubricated surfaces. As both, the cloak and the wetting ridge originate from thermodynamic forces, avoiding their formation will be challenging. We still lack a quantitative understanding of how much lubricant is depleted per drop which slides off a LIS.\\
    The prevalence of LIS in the domain of self-cleaning, shear-reduction, and biocompatible surfaces has shown great promise for use in consumable and biotechnological applications. The growth and incubation of cells in LIS surfaces could potentially provide microenvironments for single cell sorting and growth mediums. The unique hierarchical structures of LIS can provide high interfacial interactions for applications requiring cell growth, such as oxygenation or growth media replenishment.
    
    \item \textit{Elastomers} - The elastomeric ridge is arguably the most complex, owing to the coupling between the liquid and crosslinked PDMS. No other PDMS surface possesses the same degrees of freedom in tunability: crosslink density and distribution of free chains, shear modulus, oil viscosity, oil chemistry, oil-to-network affinity, oil surface tension, and polymer network surface energy. We only recently started to understand the process of oil separation in the wetting ridge, but the consequences of and for sliding drops are still largely unexplored. This is especially evident from a theoretical point of view. What is the interplay between the oil and the crosslinked PDMS in the wetting ridge and how does it change during dynamic wetting? Even for a static scenario, it is still partially unclear what causes separation, or more importantly, what keeps the lubricant trapped inside the network. Answering these question is a key aspect of designing long-term stable elastomeric coatings.\\
    The formation of cloaks and accumulation of ridges that are composed of PDMS oil remains one of the most severe degradation issues of PDMS-based surfaces. Swelling of PDMS elastomers with liquids that show much less interactions with wetting liquids (e.g. fluorinated oils) compared to PDMS oils (high spreading coefficients) may resolve or slow down the engineering challenge of separation, cloaking, and gradual surface degradation. Further investigation into the use of sustainable infusion liquids or network materials could help facilitate the future development of long-term functional and naturally degradable materials.
    
    \item \textit{SOCALS} - Wetting ridges on SOCAL are inherently hard to image and investigate, due to the nanometric length scale. While state-of-the-art computation methods are providing a greater understanding of SOCAL surfaces, experimental validation is needed to validate the accuracy of these computational models. This is particularly challenging for water drops. Detailed experimental investigations suffer from the combination of the nanometric length scales and high surface curvatures around the three-phase contact line, caused by contact angles close to 90$^\circ$. To obtain the required resolution, devices with higher sensitivity need to be developed. Alternatively, the problem can be approached using drops with lower liquid polarity, circumventing the challenges accompanied by high curvatures and contact angles. For nonpolar drops, nanometric analytical techniques such as interferometric microscopy (cf. Appendix – Methods and Techniques) might provide sufficient in-depth resolution to resolve the dynamic reorganization of the wetting ridge. A key goal would be to develop constitutive models to describe dissipation on SOCAL surfaces.\\
    Tuning the silane chemistry on SOCAL surfaces may provide gradients or patterns in glass transition properties, thereby enabling temperature-sensitive surfaces. The use of PDMS may also help shield defects of other organic-material based surfaces owing to the “self-smoothing” nature of highly flexible siloxane chains. Co-integrating different surface chemistries may optimize chain flexibility, giving rise to multi-functional surfaces capable of next-generation easy-to-clean surfaces. Promising applications include on-land or floating photovoltaic applications. More intensive studies are needed to investigate their potential in particulate or dust removal while maintaining high energy throughput during operation.
\end{itemize}

We are currently observing a transition in the perspective of soft wetting research, shifting from a fundamental understanding to a focus on practical applications, encompassing fields such as energy and biomedical sciences. Overall, PDMS-based surfaces show superior UV-stability, combined with easy-to-clean performance. The integration of PDMS-based surfaces with other organic or inorganic materials may realize useful engineering concepts. For instance, in the emerging field of synthetic biology, PDMS-based microfluidic devices play a crucial role for the synthesis of a broad range of artificial cell precursors. The subsequent assembly of these complex building blocks in functional units remains an ambitious goal of synthetic chemists, physicists and biologists. Fully exploring the potential of PDMS-based surfaces may represent a pathway to improve the synthesis and assembly of cellular building blocks (membranes, proteins, cytoskeleton, etc). Furthermore, the surface energy of PDMS is almost as low as that of environmentally harmful perfluoroalkyls. It may therefore serve as a viable replacement material for many electrode (i.e., Nafion) or membrane materials (i.e. PTFE-based) that are at the core of our energy-water infrastructure.\\
To utilize soft wetting effectively in any application, fundamental understanding is still critical, as illustrated in the self-cleaning context (cf. Coating Comparison). This review serves to provide a series of guidelines that link the current state-of-the-art soft wetting knowledge - in particular on silicone surfaces - to mechanistic features desirable for a myriad of applications.

\section*{Conflicts of interest}

The authors have no conflicts to disclose.

\section*{Acknowledgements}
We wish to thank H.-J. Butt, Z. Cai, S. Karpitschka, F. Schmid, and P. Stephan for stimulating and helpful discussions. L.H. thanks T. Boeddeker and K. Sporbeck for feedback on the manuscript. This research was motivated by the DFG Priority Programme 2171. Financial support came from the DFG Emmy Noether Programme No. 460056461 (L.H.), the EPSRC National Fellowship in Fluid Dynamics Grant No. EP-X028410-1 (A.N.), the European Union's HORIZON research and innovation program under the Marie Sklodowska-Curie grant agreement No. 101062409 and the Academy of Finland grant agreement No. 13347247 (W.S.Y.W), the US National Science Foundation through CBET grant 2326933 (J.T.P.), and the Max Planck Center for Complex Fluid Dynamics - University of Twente (D.V.).

\section*{Appendix - Methods and Techniques}

%{\color{red} 1. Name of the technique and general function; 2. What is it used for for PDMS/wetting; 3. How the technique works; 4. Strong points of the technique (e.g. confocal is good for imaging vertical slices and interfaces between two liquids)?; 5. Limitations of the technique (e.g. confocal cannot be used with non-transparent specimens).}

\subsection*{Optical Characterization}
\subsubsection*{Shadowgraphy}
The Shadowgraphy method captures shadows of macroscopic objects (e.g., drops of $>1~\mathrm{\mu l}$, cf. Fig. \ref{fig:pdms_based_coatings}, \ref{fig:optical_techniques}) with high-resolution optical cameras and back-illumination. Shadowgraphs are typically well-contrasted and thus suitable for contour profile extractions of static/dynamic drops using image processing. For example, shadowgraphs are the basis of goniometry, \textit{i.e.}, one of the most common and simplest methods to measure surface wettability ($\theta$, $\theta_a$, $\theta_r$, and $\Delta \theta$). Here, the wetting contact angle is extracted from the drop contour and recovered by various fitting techniques ( e.g. Young-Laplace equation, spherical/elliptical fit, or tangent fit \cite{kalantarian_methodology_2009, yuan_contact_2013}). Goniometry was utilized to demonstrate surface adaptation and dissipation on PDMS gels \cite{wong_adaptive_2020, karpitschka_droplets_2015}. The overall shape of the drop contour is governed by the surface tension. If the drop becomes larger than the capillary length, gravity distorts the contour shape from a perfect (hemi)sphere (i.e., oblate sessile drops and prolate pendant drops). When the density of the drop and the prevailing gravity is known, the surface tension ($\gamma$) can be measured from shadowgraphs (cf. pendant drop measurement). With optically-assisted magnification, microscopic features such as wetting ridges become visually detectable \cite{vangorcum_spreading_2020}. In place of (visible) light, x-rays were used for high-definition imaging of static wetting ridges (cf. $\lambda_e$) \cite{park_visualization_2014}, Fig \ref{fig:wetting_ridge}b. 

\subsection*{Schlieren}
Shadowgraphy setups can be upgraded to perform quantitative Schlieren images \cite{settles_schlieren_2001}. For example, wetting-induced deformations of thin PDMS elastomers were measured with Schlieren setups \cite{zhao_geometrical_2018}. For this, a collimation lens aligns the illuminating light directly after its source. The parallel light travels through the specimen (i.e., the PDMS surface) and is subsequently focused by a condenser lens. Placing a blade edge at the focal point of the condenser lens blocks the unperturbed light beams. However, any slopes on the surface of the PDMS specimen refract the light and render a visible Schlieren image behind the blade edge. The image intensity of the refracted light and the slope angle of the surface are related and can be found via appropriate calibration.

\subsubsection*{Confocal Microscopy}
Confocal microscopy is a non-invasive optical technique that allows us to image samples across both horizontal and vertical planes. Three-dimensional images can be reconstructed by stacking the two-dimensional horizontal/vertical images. Confocal microscopy has been used to study the static and dynamic properties of drops on liquid-infused surfaces and soft elastomeric surfaces (e.g., $\lambda_l$ and $\lambda_e$)\cite{jerison_deformation_2011,schellenberger_direct_2015,cai_fluid_2021,hauer_phase_2023},  cf. Fig. \ref{fig:wetting_ridge}a, \ref{fig:oil_cloak_lscm}ab, \ref{fig:wetting_ridge_dyn}b inset. Fundamentally, confocal microscopy is an extension of fluorescence microscopy where light that is out of focus is filtered out using a pinhole placed at the conjugated point of the optical path. Laser scanning confocal microscopes scan across a sample point by point and combine the scans to produce 3D images with a horizontal resolution of around 200 nm and a vertical resolution of around 1 $\mu$m, Fig. \ref{fig:optical_techniques}. The relatively slow point-by-point scan can be accelerated with a spinning archimedean disk. When using confocal microscopy, different fluorescent dyes must be added to the different phases (e.g. drop and coating) being imaged. The use of dyes enables the detection of interfaces between phases of similar refractive index. Typically, inverted confocal microscopes (objective lens below specimen) are used to image drops because this configuration makes it possible to image the contact region between the drop and the surface - which is the region that governs the wetting properties - with minimal optical artifacts. The primary limitation of this technique is that it requires transparent substrates, such as glass.

\subsubsection*{Interferometry}
\begin{figure}[h!]
    %\centering
    \includegraphics[width=0.5\textwidth]{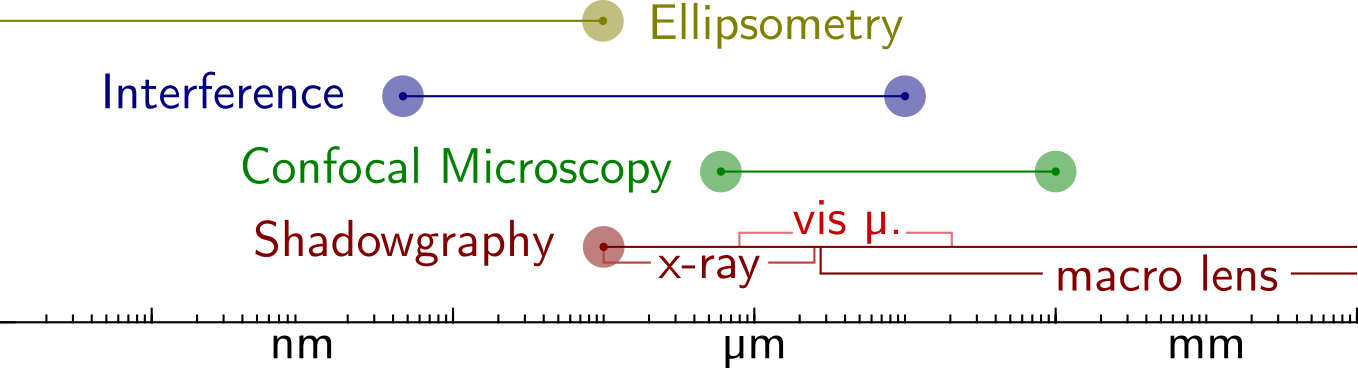}
    \caption{\label{fig:optical_techniques} Different imaging techniques are suitable for different length scales. Most light-based techniques such as (visible) shadowgraphy or confocal microscopy are bound by the diffraction limit. Interferometry and ellipsometry overcome this limit by exploiting wave superposition and phase shifts, bringing down the resolution limit to sub-nano.}
\end{figure}

Interferometric microscopy enables in-plane depth measurements with nano-to-microscopic resolution. In wetting, this has been utilized to directly visualize nanometric features such as contact angles/lines ($\theta$) \cite{shoji_measurement_2021,papadopoulos_longterm_2016}, drop profiles \cite{li_evaporating_2020}, wetting ridges ($\lambda_l$ and $\lambda_e$) \cite{carre_viscoelastic_1996, mitra_probing_2022}, and thin films \cite{liu_coalescence_2019,vakarelski_interferometry_2022,deruiter_wettabilityindependent_2015,daniel_oleoplaning_2017, panchanathan_levitation_2021}. Coherent light of a “diagnostic beam” travels through the specimen and is brought into superposition with a reference beam with identical coherency. The emerging interference pattern gives the optical path (geometric depth times refractive index) difference between both light beams. The optical path length of the reference beam is calibrated. Different ways of beam calibration brought up various flavors of interference techniques such as Reflection Interference Contrast Microscopy, Digital Holographic Microscopy, or Michelson interferometry, each varying in practical complexity and precision. With information on the refractive index of the specimen, the geometrical depth of the specimen can be determined with a precision of $\le 10~\mathrm{nm}$, Fig. \ref{fig:optical_techniques}. 

\subsubsection*{Ellipsometry}
Ellipsometry is a non-destructive/non-invasive optical technique that allows the characterization of thin films with sub-Angstrom resolution. The fast scanning speed ($<1$ min for single-wavelength, $<1$ s for multi-wavelength) enables in situ measurements of dynamic films. The technique was used to investigate e.g., precursor spreading of PDMS oil on glass \cite{beaglehole_profiles_1989,heslot_dynamics_1989,heslot_experiments_1990} or wetting on SOCAL surfaces ($H_B$) \cite{villette_ultrathin_1997, lhermerout_contact_2019}. Ellipsometry measures changes in the polarization state of an incident light beam upon interaction with the surface. In ellipsometric measurements, linear polarized light is directed to a film. Depending on the thickness and material properties of the film, the amplitude and phase of the reflected light change due to light-film interactions (absorption/scatter). The light-film interactions are different for the p and s components of the light (parallel and lateral to the plane of the incident), resulting in an altered, elliptical polarization state of the reflected light. The ratio of s/p amplitudes before and after reflection, and the phase shift contains information on the film thickness and optical constants (e.g., refractive index). Ellipsometers either use a single wavelength or multiple wavelengths. In the former case, the angle of incidence is varied and the layer thickness and refractive index are fitted. When working with multiple wavelengths, usually one angle of incidence is used and the film thickness and refractive index are fitted as a function of wavelength. PDMS films on oxidized surfaces (\textit{i.e.}, glass or silicon oxide) are prone to measurement errors as both have almost identical refractive indices. 

\subsection*{Mechanical Characterization}

\subsubsection*{Classical Rheology} 
Rheology describes the stress-strain/shear behavior and the mechanical properties of soft materials such as PDMS. From a wetting perspective, these mechanical properties provide information related to the dynamics and dissipation in the wetting ridge and around the contact line. The term rheology is generic, which includes various geometries and experimental setups to measure dynamic mechanical properties. The most common for silicone-based materials - particularly related to wetting - is parallel-plate shear rheology. In these experiments, silicone-based elastomers, swollen networks, or melts are loaded between two plates. The top plate is typically rotated in a sinusoidal fashion while recording both the stress and the strains. The lower plate can be connected to a Peltier plate, or the entire system can be enclosed into an environmental chamber if temperature or humidity control is needed. Typically, an amplitude strain sweep is conducted at a constant frequency to determine the linear region. A constant strain is then chosen, and a frequency sweep is then conducted to quantify the mechanical properties of the material as a function of frequency (i.e., rate dependence) in the linear regime. Common values that are extracted from the measurement as a function of frequency are the storage $(G')$ and loss modulus ($G''$). For more liquid-based PDMS (i.e. oils), viscosities can also be measured through a similar method. The storage modulus describes the tendency of the material to store energy (solid-like character) whereas the loss modulus is the tendency to dissipate energy (liquid-like character). Their ratio, $G''/G'=\tan \delta$ can be associated with damping. 

\subsubsection*{Indenter} 
The use of indentation and contact mechanics is useful for characterizing surface and near-surface properties. This can include properties like adhesion and friction, lubricant layer properties, and the mechanical properties of a substrate. Such experiments can be conducted across various size scales, from macro- \cite{darby_modulus_2022,dorogin_contact_2018,shull_contact_2002} to micro- \cite{bowen_application_2011,glover_creasing_2023} to nanoscales \cite{wang_nanoindentation_2015}. The experimental setup generally consists of a spherical or cone shaped probe, which is attached to a force measurement tool (e.g. a deflecting cantilever or a load cell). For the nanoscale, an AFM tip is usually employed. In a typical experiment, the probe approaches the surface at a designated rate until it comes into contact, at which point the probe and surface can jump into contact. The probe continues to indent into the surface until reaching a pre-defined distance or force. The probe is then either retracted from the surface directly or held stationary for some time and then retracted. During this entire process, the force and distance are recorded, allowing for quantification of mechanical properties during the approaching process, relaxation of the material during the holding time, and adhesion during the retraction process. Combining indentation with contact mechanics models can enable the quantification of mechanical properties while integrating force-distance curves during retraction determines the work required to separate two surfaces. By sliding laterally, friction measurements can be conducted with a similar type of experimental setup. From the wetting perspective, indentation can be used as a tool to characterize mechanical characteristics, which can then be related to the wetting ridge formation. For example, measuring forces during the holding time can provide a route to separate the viscoelastic and poroelastic relaxation timescales associated with wetting ridges \cite{hu_indentation_2011}. Moreover, particle indentation methods can also be connected to fluid separation and surface stress. The indentation process can also provide information for pure liquid surfaces like capillary forces \cite{rahat_capillary_2023}, detection of nanobubbles \cite{peppou-chapman_detection_2022} or quantification of lubricant layer thicknesses \cite{peppou-chapman_mapping_2018} for SLIPS. 

\subsubsection*{Force Microscopy} 
\begin{figure}[h!]
    %\centering
    \includegraphics[width=0.5\textwidth]{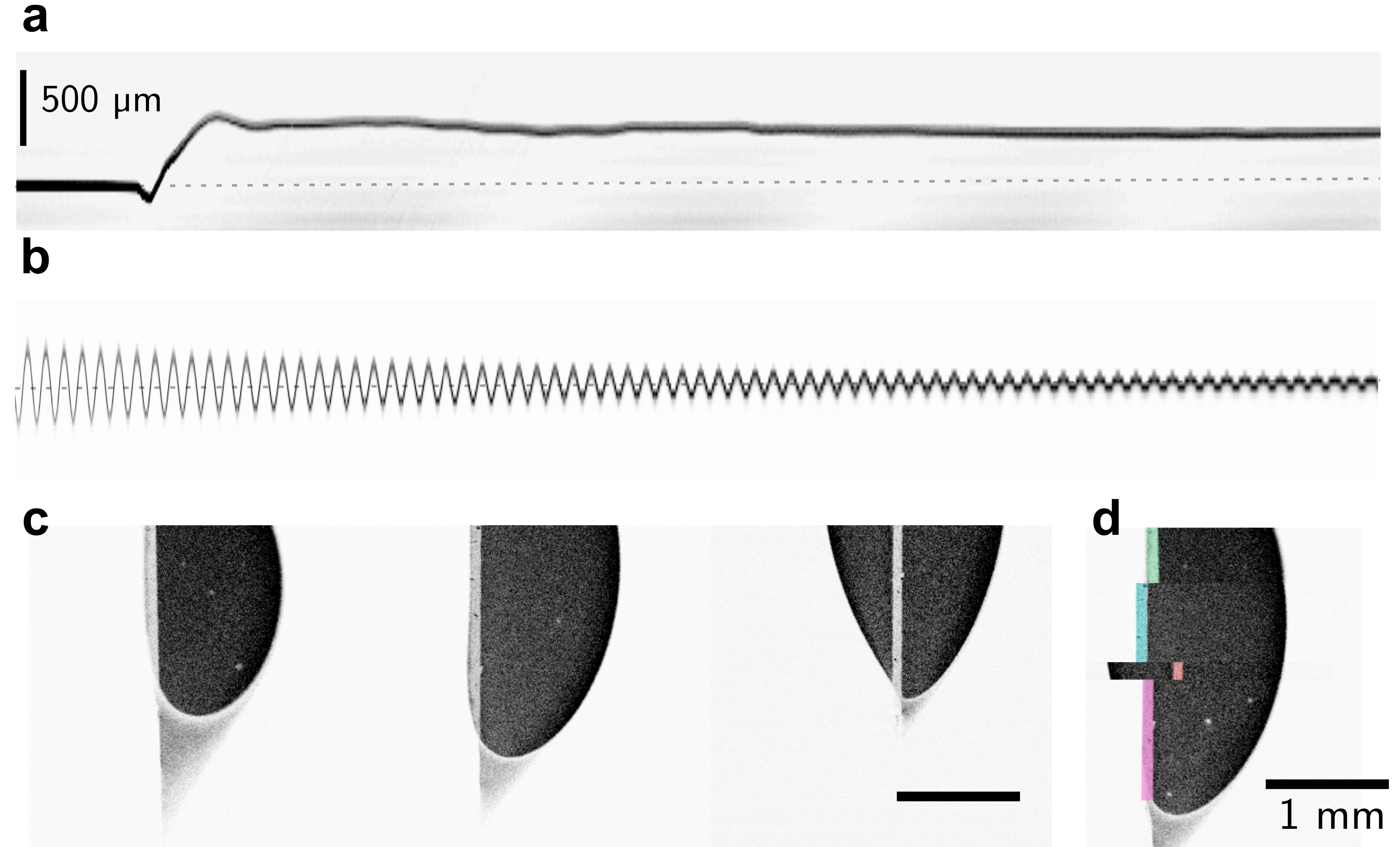}
    \caption{\label{fig:force_miscroscope} Force microscopic measurement conducted on a confocal microscope. Here, $10~\mathrm{\mu l}$ water drops slide on on SOCAL surfaces (cf. Fig. \ref{fig:wetting_ridge_dissipation}c). a) Kymograph of the blade deflection upon drop contact, measured with the reflection channel of an $xt$-scan. b) Calibration curve to determine the spring constant of the blade, here approx. $0.21~\mathrm{N/m}$. c) Atto-dyed, steady-state drop footprints, and faintly the probing metal blade. d) overlay of various relaxed drop footprints.}
\end{figure}

Forces that build up during drop-sliding on PDMS films can be measured with cantilever set-ups \cite{pilat_dynamic_2012}. Such “force microscopes” were used to measure dissipation on PDMS surfaces of all types, Fig. \ref{fig:wetting_ridge_dissipation}. The cantilever is vertically placed over the PDMS surface. Glass capillaries, metal blades, and rods with ring ends have been successfully utilized as cantilevers \cite{pilat_dynamic_2012, gao_how_2018, beitollahpoor_determination_2022, khattak_direct_2022, hinduja_scanning_2022}. The cantilever top is fixed while the lower end hangs freely several micrometers above the surface. Displacements of the lower end (Fig. \ref{fig:force_miscroscope}a) - which can be tracked with e.g. shadowgraphy - relate to acting forces and can be extracted with the cantilever spring constant. The spring constant is calibrated by measuring the natural frequency (Fig. \ref{fig:force_miscroscope}b) or by applying well-defined loads. To measure dynamic wetting forces on PDMS surfaces, a sessile drop is pushed against the cantilever by moving the surface horizontally. The major drawback of this technique is the invasiveness of the capillary to the drop, \ref{fig:force_miscroscope}c. It is still not completely understood how the cantilever itself influences the measured friction forces.

\subsection*{Computational Methods}

\subsubsection*{Molecular Dynamics (MD) Simulations}

Molecular dynamics (MD) simulations are particle-based methods that resolve the evolution of the particle positions and momenta by solving Newton's second law of motion \cite{frenkel2023understanding}. Due to the molecular resolution, Fig. \ref{fig:comput_techniques}, MD is utilized to obtain explicit insight into the polymer behavior of PDMS chains during wetting. In particular, MD is used to investigate the nanoscopic wetting ridges of SOCAL surfaces, and the conformation/dynamics of PDMS chains, or to calculate relevant thermodynamic quantities such as surface tensions or micro-rheology. The particles can represent atoms but also larger, coarse-grained structures (e.g. molecules), which affects the interaction force fields between particles. Interaction forces may be atomistic (quantum mechanical) or effective (e.g. Lennard-Jones potential) such that they reproduce the most important physical properties of the system of interest. By virtue of their resolution, MD simulations are computationally expensive since one needs to solve the equation of motion for every particle. Consequently, MD setups are usually microscopic in time and space resolution.

\subsubsection*{Lattice Boltzmann Method (LBM)}

\begin{figure}[t]
    %\centering
    \includegraphics[width=0.5\textwidth]{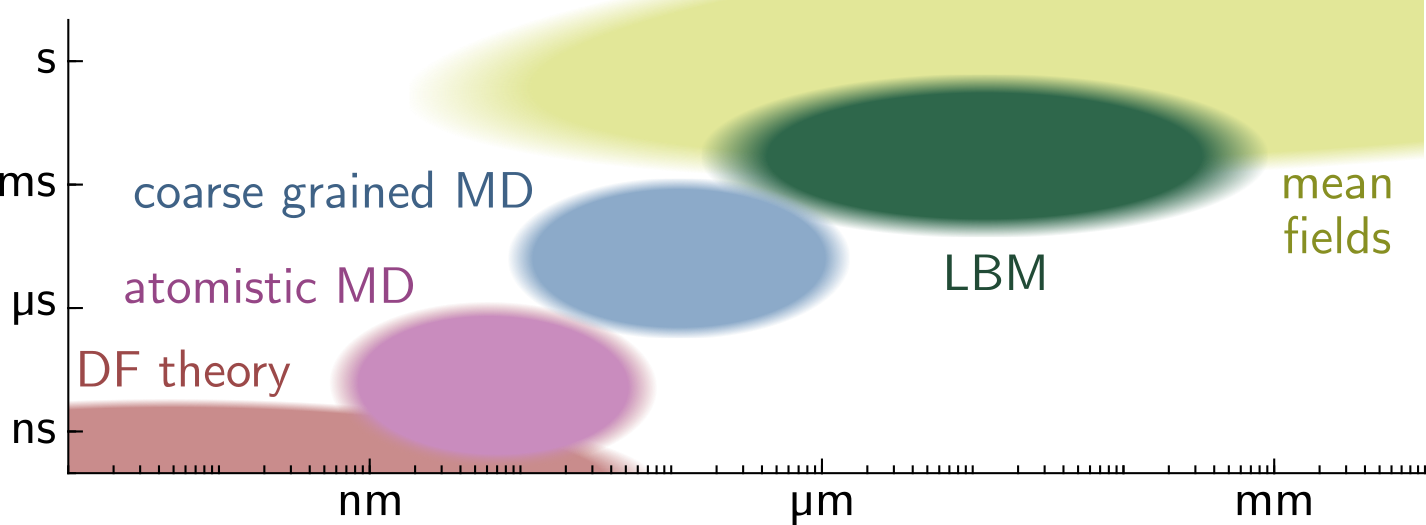}
    \caption{\label{fig:comput_techniques} Typical time and length scale domains of wetting problems, treated with different computational methods. Density functional theory (DFT) is utilized in the sub-nano domain. The next larger domain sizes are treated with atomistic and coarse-grained molecular dynamics (MD) simulation. The Lattice Boltzmann method (LBM) utilizes kinetic theory and is mesoscopic, linking the microscopic and the macroscopic world. Mean field theory is utilized for large molecule ensembles and macroscopic problems.}
\end{figure}

 LBM is a computational method typically applied to solve fluid mechanics problems \cite{kruger_lattice_2017,haghanihassanabadi_numerical_2018,mazloomim_entropic_2015}. In contrast to microscopic methods (e.g. MD) that compute the motion of individual (or clusters) of fluid particles, LBM is a mesoscopic method based on kinetic theory, Fig. \ref{fig:comput_techniques}. The central quantity computed in LBM is the probability distribution function, which gives the number of fluid particles moving at a given velocity at a given position. From the distribution function, we can obtain detailed information that is difficult to obtain experimentally, such as the velocity field, pressure distribution, and viscous dissipation in the fluid. LBM has been applied to study drop dynamics on flat and textured surfaces, including liquid-infused surfaces \cite{sadullah_drop_2018}. One of the key advantages of LBM is that coarse-grained molecular interactions (for example, between liquids and solid surfaces) and complex solid geometries can be implemented efficiently. Since LBM focuses on the collective behavior of fluid particles rather than on individual fluid particles, it is less computationally intensive than microscopic simulations, which means larger drops can be simulated relative to the surface features (height of pillars in the case of liquid-infused surfaces).

\subsubsection*{Mean Field Models}

Mean field approaches regard the collective behavior of fluids and solids and are an alternative to MD and LBM, particularly for macroscopic systems, Fig. \ref{fig:comput_techniques}. Mean field models were utilized e.g., for static and dynamic wetting on PDMS gels \cite{karpitschka_droplets_2015, brummelen_binaryfluid_2017, thiele_gradient_2020, aland_unified_2021, henkel_soft_2022}, characterization of wetting ridges \cite{rivetti_elastocapillary_2017, pandey_singular_2020, flapper_reversal_2023, gupta_numerical_2023}, water sorption in PDMS \cite{baumli_linear_2022}, or to model cloaking \cite{islam_numerical_2023}. Mean field models exploit that individual fluctuations of atoms, molecules, or coarse-grained particles diminish when averaged over sufficiently large ensembles. As a result, energy and momentum are not properties of single beads, but (weakly) smooth mean fields which are time and space-dependent. The variety of mean-field approaches is large: Momentum-based approaches applied to fluids result in the well-known Navier-Stokes equations and their many sub-frameworks (lubrication theory, Stokes equation, potential theory) \cite{slattery_advanced_1999, acheson_elementary_1990, probstein_physicochemical_2005}. Solids are frequently described with linear elasticity theory. Complex materials require rheological models (e.g. Maxwell/Kelvin-Voigt bodies). Consideration of the (free) energy field gives rise to the Allen-Cahn/Cahn-Hilliard equations \cite{allen_coherent_1975, cahn_free_1958}, gradient dynamics \cite{li_unified_2022}, or the heat equation \cite{einstein_ueber_1905}. The spatiotemporal distribution of the mean-field quantities is expressed with partial differential equations (PDE). Finding closed-form solutions for these PDEs is the exception as the dependent variables (e.g., velocity and temperature) can be coupled and the system geometry is often complex. Hence, numerical tools, such as the Crank-Nichelson method and the finite element method are utilized to solve the time and space components, respectively \cite{eriksson_computational_1996, ferziger_computational_2002, hirsch_numerical_2007}. Multiple phases and their shared interfaces are handled with explicit (interface tracking, e.g., arbitrary Lagrangian-Euler \cite{hirt_arbitrary_1974}) or implicit (interface capturing, e.g. volume of fluid \cite{hirt_volume_1981, osher_fronts_1988}) methods. Commercial and open-source software libraries simplify the practical application of such tools. Mean field models assume material properties (e.g., densities, viscosities, compliances, mobilities, surface tensions, etc.) \textit{a priori}, which can be a drawback for various PDMS wetting-related questions (e.g. adaptation or cloaking). Also, the individual molecular behavior of PDMS molecules is inaccessible.


%
% ****** End of file apssamp.tex ******

%\nocite{*}

%\bibliographystyle{apsrev.bst}
\bibliography{rsc}% Produces the bibliography via BibTeX.

\end{document}